\documentclass{elsarticle}
\usepackage{amssymb}
\usepackage{color}
\journal{Earth and Planetary Science Letters}

\newcommand{\mib}[1]{\mbox{\boldmath$#1$}}
\newcommand{\tx}[1]{\textrm{\scriptsize #1}}

\newcommand{\revone}[1]{\textcolor{black}{#1}}  %for Reviewer 1 at Revision 1
\newcommand{\revtwo}[1]{\textcolor{black}{#1}}  %for Reviewer 2 at Revision 1
\newcommand{\revoneR}[1]{\textcolor{black}{#1}}   %for Reviewer 1 at Revision 2

\begin{document}

\begin{frontmatter}

\title{Anelastic torsional oscillations in Jupiter's metallic hydrogen region}

%\author{K.~Hori\fnref{myfootnote}, R.J.~Teed, C.A.~Jones}

\author[mymainaddress,mysecondaryaddress]{K.~Hori\corref{mycorrespondingauthor}}
\cortext[mycorrespondingauthor]{Corresponding author}
\ead{amtkh@leeds.ac.uk}

\author[mythirdaddress]{R.J.~Teed}
\author[mysecondaryaddress]{C.A.~Jones}

\address[mymainaddress]{Graduate School of System Informatics, Kobe
 University, Kobe, Japan}
\address[mysecondaryaddress]{Department of Applied Mathematics, University of Leeds, Leeds, UK}
\address[mythirdaddress]{School of Mathematics and Statistics, University of Glasgow, Glasgow, UK}

\begin{abstract}
%This template helps you to create a properly formatted \LaTeX\ manuscript.
%
%
We consider torsional Alfv\'{e}n waves which may be excited in Jupiter's
 metallic hydrogen region.
\revone{\revoneR{These} axisymmetric zonal flow fluctuations have previously been
 examined for incompressible fluids
 in the context of Earth's liquid iron core.}
\revoneR{Theoretical models of the deep-seated Jovian dynamo,
 implementing radial changes of density and electrical conductivity
 in the equilibrium model},
 have reproduced its strong, dipolar magnetic field.
\revoneR{Analyzing such models, we find anelastic torsional waves travelling perpendicular to the rotation axis in the metallic region on timescales of at least several years. 
Being excited by the more vigorous convection in the outer part of the dynamo region, they can propagate both inwards and outwards.
When being reflected at a magnetic tangent cylinder at the transition to the molecular region, they can form standing waves.
Identifying such reflections in \revoneR{observational} data could determine the depth \revoneR{at which} the metallic region effectively begins.}
Also, this may distinguish Jovian torsional waves from those in Earth's core,
 where observational evidence has suggested waves mainly travelling outwards
 \revone{from the rotation axis}.
\revoneR{{These waves can transport angular momentum}
 and possibly give rise to variations in Jupiter's rotation period of
 {magnitude no greater than tens \revoneR{of}} milliseconds.}
In addition
 \revoneR{\revoneR{these}  internal disturbances could \revoneR{give rise to a 10\% change over time in the zonal flows 
at a depth of 3000\,km below the surface.}}
\end{abstract}

\begin{keyword}
%\texttt{elsarticle.cls}\sep \LaTeX\sep Elsevier \sep template
%\MSC[2010] 00-01\sep  99-00
Waves \sep Jupiter \sep Magnetic field \sep Interior \sep Zonal jets \sep Length of day
\end{keyword}

\end{frontmatter}

%\linenumbers

\section{Introduction}  \label{sec:intro}

Torsional Alfv\'{e}n waves (TWs) are a special class of
 magnetohydrodynamic (MHD) \revoneR{waves
 whose} transverse motions are confined to cylindrical surfaces aligned with the rotation axis.
They are perturbations about the Taylor state \citep{Tay63} expected at leading order
 when the Coriolis, Lorentz, and pressure gradient forces are in balance
 in the momentum equation, the so-called magnetostrophic balance.
The linear theory for incompressible Boussinesq fluids was
 introduced by \citet{Bra70}
 and has been applied to Earth's core,
 in which fluid motions of liquid iron are believed to
 generate the global, intrinsic magnetic field. 
The axisymmetric disturbances can propagate in cylindrical radius
 \revone{(denoted by $s$ hereafter)}, perpendicular to the rotation axis.
This enables the waves
\revone{to transport the angular momentum
 to other regions}, including the rocky mantle and solid inner core, 
 through electromagnetic, gravitational, topographic, and viscous
 couplings (see \citep{RA12} for an overview).
The evidence for such waves within the {Earth's} fluid core
 has been discussed using 
 core flow models inverted from the observed geomagnetic secular variation (SV):
 the zonal component was found to exhibit
 fluctuations with a near six-year period \citep{GJCF10,GJF15}.
They may also account for a decadal variation of the length-of-day (LOD)
 of Earth \citep{HV13}.
Such information provides insight on the deep interior
 by constraining physical quantities that cannot be measured directly,
 such as the field strength within the core
 and the electrical conductance of the lowermost mantle.

 Here we extend the study of TWs to compressible fluids by applying
 the anelastic approximation, {in which} sound waves are ignored.
This is of some interest for geophysical modelling,
 since there is a density increase of 22\% from the bottom to the top
 of Earth's fluid outer core \citep[e.g.][]{JF15}.
{The} extension to anelasticity is, however, more strongly motivated by a desire to explore the internal
 dynamics of gas planets and stars,
 which mostly consist of hydrogen and helium.
Jupiter is the largest planet in the solar system
 and also has the strongest planetary magnetic field, with a surface magnitude of 
 $\sim 10\,\textrm{G}$, or $1\,\textrm{mT}$. 
Dynamo action is predicted to operate in a metallic hydrogen region
  situated below a molecular hydrogen envelope. 
The phase transition is expected to occur continuously
\revtwo{between 0.85 and 0.90\,$R_\tx{J}$,}
 with $R_\tx{J}$ being Jupiter's mean radius at the 1 bar level.
Adopting an interior model including the transition \citep{FBLNBWR12},
 dynamo simulations for anelastic convection have 
 reproduced Jupiter-like magnetic fields \citep{J14,GWDHB14}. 
\makeatletter
The gas giant is rapidly rotating with a period of 9.925 hours (the System \@Roman{3});
\makeatother
 changes on the order of tens of milliseconds have been noted \citep{HCR96,RYK01}.
\revoneR{The rapid rotation and strong magnetic field in the metallic region
    give rise to a force balance in which the viscous forces are small
compared to the Coriolis and Lorentz forces: the quasi-magnetostrophic balance. 
Jupiter may therefore be a good candidate for hosting TWs.}

\revone{MHD waves excited within the gas giant may produce decadal variations,
 as shown below.}
In-situ observations of Jupiter have the longest history amongst all
 planets other than Earth,
 spanning over forty years since the Pioneer epoch in the early 1970s. 
Coverage is however sparse; although
data retrieved from past missions \revoneR{have} enabled the construction of
 global models for the magnetic field, such data was limited to
 spherical harmonics of degree no higher than seven \citep{C93,RH16}.
\citet{RH16} showed time-dependent field models to be preferable to steady models,
 and attempted to invert the SV to flows 
 at the top of the expected metallic region.
The Juno spacecraft is currently orbiting the gas giant and
 the newly available data 
\revoneR{sample the field closer to its source than for any other planetary dynamo so far} \citep{Betal17,Betal17b,JH17,C18}.
Over the planned five-year mission
 it will better define the field in both temporal and spatial resolution.
Also, the gravitational sounding has indicated that
 zonal flows extend thousands of kilometres below the Jovian surface
 \citep{KGHetal18,GMMetal18}.
The cloud motion has been tracked for decades by Earth-based telescopes
 to measure changes to, or periodicities in, the zonal winds \citep{TWPSetal17}.
The colouration, brightening, and outbreak events,
 sometimes leading to global upheavals, have been
 \revone{monitored}
 for more than one hundred years \citep{R95,F17}.

Of more theoretical interest is the nature of excited TWs.
Since the Alfv\'{e}n \revone{waves} are able to propagate in
 \revone{$s$} inwardly and outwardly,
 early studies proposed TWs in the form of standing waves and sought
 wave motions in the form of normal mode solutions \citep{Bra70,ZB97,BMJ09},
 often referred to as torsional \emph{oscillations}.
However, Earth's core flow inversions/assimilation \citep{GJF15}
 and numerical geodynamo simulations \citep{WC10,TJT14,SJNF17}
 have \revoneR{found} TWs travelling predominantly in an outwards direction
 with no obvious reflection at the boundaries.
This could be explained through preferred excitation \citep{TJT14,TJT15,TJT19}
 near the tangent cylinder (TC, the imaginary cylinder aligned with the
 rotation axis that circumscribes the inner core)
 \revone{and dissipation beneath and above the core-mantle boundary (CMB)
 \citep{SJCD12,SJ16}}.
Studies ignoring dissipation showed that 
 the effect of spherical geometry and variable internal magnetic fields
 can give rise to asymmetric reflections and hence weaken reflected waves
 \citep{CLM14}.
We shall demonstrate that TWs in the gas giant's metallic region
 can be reflected from a magnetic TC,
 which is formed due to the transition to molecular hydrogen. \revoneR{This leads 
to the formation of standing torsional waves}.
% its standing nature.

\section{Theory} \label{sec:theory}

The theoretical framework of incompressible TWs is well documented
 \citep[e.g.][]{Bra70,TJT14,JF15}.
In the light of those studies, \revone{we consider anelastic fluids
 where the Lantz-Braginsky-Roberts formulation \citep{BR95,LF99,J11} is adopted
 and explore anelastic TWs within the electrically conducting region of the gas planet}.
\revone{We assume
 the equilibrium state is close to adiabatic, well-mixed, and hydrostatic with density $\rho_\tx{eq}$.
The velocity perturbations of the waves \mib{u} are subsonic, 
 so that the continuity equation is}
 \begin{equation}
  \nabla \cdot \rho_\tx{eq} \mib{u} = 0  \; .\label{eq:continuity}
 \end{equation}
We assume a basic state dependent only on spherical radius, $r$,
 and denote it by subscript 'eq' hereafter.
We focus on the  rapid dynamics,
 in which the characteristic timescale is much shorter than the diffusion time.
This allows us to begin with the momentum equation
\begin{equation}
  \frac{\partial \mib{u}}{\partial t} + (\mib{u} \cdot \nabla ) \mib{u}
 + 2 \mib{\Omega} \times \mib{u} 
  = -\nabla \hat{p}
   + \frac{1}{\rho_\tx{eq}} \mib{j} \times \mib{B}
   - \hat{\mib{e}}_r \frac{d T_\tx{eq}}{dr} S  \; ,
\end{equation}
 where $\mib{\Omega}$ is the rotational angular velocity, 
 $\mib{j}$ is the current density, $\mib{B}$ is the magnetic field,
 $S$ is the entropy, $T_\tx{eq}$ is the temperature in the
 equilibrium state, $\hat{\mib{e}}_r$ is the unit vector in the
 radial direction,
 and $\hat{p}$ is a reduced pressure incorporating the density
 and the gravitational potential. 
Hereafter we suppose $\mib{\Omega} = \Omega \hat{\mib{e}}_z$
 with $ \hat{\mib{e}}_z$ being the unit vector in the direction of rotation axis.
To look at fluctuations corresponding to TWs,
 we consider the axisymmetric $z$-independent azimuthal flow
 by taking averages of the $\phi$-component of the momentum equation
 over cylindrical surfaces to give 
\begin{eqnarray}
 \frac{\partial}{\partial t} \langle \rho_\tx{eq} \overline{u_\phi} \rangle
 &=& - \left\langle \overline{\hat{\mib{e}}_\phi \cdot ( \nabla \cdot
         \rho_\tx{eq} \mib{u} \mib{u}}) \right\rangle
  - 2 \Omega \langle \rho_\tx{eq} \overline{u_s} \rangle 
   + \left\langle \overline{\hat{\mib{e}}_\phi \cdot \frac{1}{\mu_0}(\nabla\times\mib{B}) \times \mib{B}}\right\rangle \nonumber \\
 &\equiv & F_\tx{R} + F_\tx{C} + F_\tx{L} \; ,   \label{eq:N-S_uphi}
\end{eqnarray}
 where the azimuthal and axial averages are defined as
 \revoneR{$\overline{f} = ({1}/{2\pi }) \int_0^{2\pi} f\, d\phi$} and
 $\langle f \rangle = ({1}/{h}) \int_{z_{-}}^{z_{+}} f\,dz$,
respectively, with $h = z_{+}-z_{-}$ for any scalar field, $f$.
\revone{Outside the TC}, the integral is limited by
 $z_\pm = \pm \sqrt{r_\tx{o}^2 - s^2}$ with $r_\tx{o}$
 being the radius of the planet.
Hereafter we shall focus on the region outside the TC.
With the divergence theorem, the Coriolis force becomes
 $F_\tx{C} = - ({\Omega}/{\pi s h}) \int \nabla\cdot {\rho}_\tx{eq}\mib{u}\,dV$
 for geostrophic cylinders.
From (\ref{eq:continuity}), this term vanishes,
 implying zero net mass flux across the surfaces.
\revone{For the magnetostrophic balance where the inertia and $F_\tx{R}$
 are negligible}, (\ref{eq:N-S_uphi}) gives $F_\tx{L} = 0$,
 i.e.~the Taylor state for anelastic fluids.
The Lorentz and Reynolds forces may be rewritten as 
 \begin{eqnarray}
 F_\tx{L} &=& 
      \frac{1}{\mu_0 s^2 h} \frac{\partial}{\partial s} s^2 h \left\langle \overline{B_s B_\phi} \right\rangle
    - \frac{1}{\mu_0 h} \left[
		  \frac{s}{z} \overline{B_s B_\phi}
		  + \overline{B_z B_\phi}
		 \right]_{z_{-}}^{z_{+}}   \nonumber\label{084216_14Dec18} \\ 
 \textrm{and} \quad 
 F_\tx{R} &=&
   - \frac{1}{s^2 h}\frac{\partial}{\partial s}
      s^2 h \left\langle {\rho}_\tx{eq} \overline{ u_s u_\phi } \right\rangle
   - \frac{1}{h} \left[
		  \frac{s}{z} {\rho}_\tx{eq} \overline{u_s u_\phi}
		  + {\rho}_\tx{eq} \overline{u_z u_\phi}
		 \right]_{z_{-}}^{z_{+}}\;,  \label{eq:FL_FR}
 \end{eqnarray}
respectively. 
\revone{The second term of each represents \revoneR{the surface term across an interface between
 the internal fluid region and the outside}, magnetically or dynamically.}
Since the currents vanish outside the metallic hydrogen zone,
 the $F_\tx{L}$ surface term will be small,
 and the average over the cylinder could be taken only
over the conducting region. 
For the stress-free outer boundary used in Jupiter simulations,
 the $F_\tx{R}$ surface term vanishes also. 
\revone{However, unlike the magnetic term, the molecular non-conducting region can contribute significantly to the $F_\tx{R}$ integral,
 because the convection-driven velocities are large there,
 as we shall see below.}

We now make the ansatz of splitting magnetic field and velocity into
 their mean and fluctuating parts: 
\begin{eqnarray}
 \mib{u} (s,\phi,z, t)
 &=& \widetilde{\mib{U}} (s,\phi,z)
   + \revone{ \langle \overline{\mib{u}'} \rangle (s,t) }
   + {\mib{u}'}^\tx{a} (s,\phi,z,t)  \nonumber \\ 
 \mib{B} (s,\phi,z,t)
 &=& \widetilde{\mib{B}} (s,\phi,z) + \mib{b}' (s,\phi,z,t) \; ,
\end{eqnarray}
where $\widetilde{f} = ({1}/{\tau}) \int f\,dt$ with $\tau$ being a time
 interval, $f' = f - \widetilde{f}$, $f^\tx{a} = f - \langle \overline{f} \rangle$,
 and $\widetilde{f'}=0$, $\langle \overline{{f}^\tx{a}} \rangle = 0$.
The time interval $\tau$ is chosen to be significantly longer than the expected wave-period, but not excessively
longer to avoid unnecessary computational expense. 
Here the induction equation for compressible fluids is
\begin{equation}
 \frac{\partial \mib{B}}{\partial t}
  = \mib{B}\cdot\nabla\mib{u} - \mib{u}\cdot\nabla\mib{B}
   - (\nabla\cdot\mib{u})\mib{B} \; . \label{eq:induction} 
\end{equation}
Recall that we primarily seek the rapid dynamics within the conducting fluid region so we ignore any dissipation; 
 the magnetic diffusion will become substantial as the wave goes up to the poorly-conducting zone, and
will damp waves through ohmic dissipation, but the frequencies and the waveforms in the  conducting
region should be relatively unaffected by diffusion.
\revtwo{We now substitute (\ref{eq:induction}) into the time-derivative of
\revone{the momentum equation} (\ref{eq:N-S_uphi}). 
There is a question of whether the TW equation should be expressed
 in terms of $\langle \rho_\tx{eq} \overline{u_\phi} \rangle/s$ or
 $\langle \overline{u_\phi} \rangle/s$, because the time-derivative of
 $\langle \rho_\tx{eq}\overline{u_\phi} \rangle$ appears in (\ref{eq:N-S_uphi}),
 but (\ref{eq:induction}) contains spatial derivatives of $u_\phi$,
 not $\rho_\tx{eq} u_\phi$. 
Here we choose $\langle \overline{u_\phi} \rangle$ as the dependent variable:
 we separate $u'_\phi$ into its geostrophic and ageostrophic parts,
\begin{equation}
 \frac{\partial^2}{\partial t^2} \langle \rho_\tx{eq}\overline{u'_\phi} \rangle  
 = \revoneR{ \frac{\partial^2}{\partial t^2} \langle \rho_\tx{eq}
    (\langle \overline{u'_\phi} \rangle +   \overline{{u'}^\tx{a}_\phi})
    \rangle } 
 = \langle \rho_\tx{eq} \rangle \frac{\partial^2 \langle \overline{u'_\phi} \rangle}{\partial t^2}
   \; + \; \left\langle \rho_\tx{eq} \frac{\partial^2 \overline{{u'}^\tx{a}_\phi}}{\partial t^2} \right\rangle \; ,  \label{eq:momentum_split}
\end{equation}
and ignore the second ageostrophic term,
 because it is small compared to the first term in our simulations.
We then obtain}
%\clearpage
\begin{eqnarray}
 \frac{\partial^2 \langle \overline{u'_\phi} \rangle}{\partial t^2}
 &=& \frac{1}{\mu_0 \langle {\rho}_\tx{eq} \rangle} \frac{1}{s^2 h} \frac{\partial}{\partial s}s^2 h
     \left\{
      \left\langle \overline{s B_s (\mib{B}\cdot\nabla) \frac{u_\phi}{s} } \right\rangle
      + \left\langle \overline{ \frac{B_\phi}{s} (\mib{B}\cdot\nabla) s u_s } \right\rangle
 \right. \nonumber \\
 &&   \vspace{30mm}
 \left. 
      - \left\langle \overline{ \left( \mib{u}\cdot\nabla + 2 \nabla\cdot\mib{u} \right) B_s B_\phi } \right\rangle
     \right\}
    +\frac{\partial}{\partial t}\frac{F'_\tx{R}}{\langle {\rho}_\tx{eq}\rangle}  \; 
 \label{eq:N-S_uphi_tderiv} 
\end{eqnarray}
where the non-fluctuating part of the first term of the right hand side
 sums up to zero, i.e. the Taylor constraint.
The theory is equivalent to that of the incompressible case \cite{TJT14}
 but now the effect of compressibility remains in the Lorentz term.
The fluctuating components are assumed to be significantly smaller than the mean parts.
\revoneR{Following section 3.1 of \cite{TJT14}, we separate the Lorentz terms in 
(\ref{eq:N-S_uphi_tderiv}) into a restoring force part $F_\tx{LR}$
and a driving part $F_\tx{LD}$, where the restoring force part is the
term coming from the axisymmetric geostrophic part of $u'_\phi$ and $F_\tx{LD}$ contains the
remaining Lorentz terms. The  $F'_\tx{R}$ part then corresponds to the driving
of the TW by the Reynolds forces. If the Lorentz and Reynolds driving terms are omitted, we obtain
the homogeneous free oscillation TW equation}
\begin{equation}
 \frac{\partial^2}{\partial t^2} \frac{\langle \overline{u'_\phi} \rangle}{s}
 = \frac{1}{s^3 h \langle{\rho}_\tx{eq} \rangle} \frac{\partial}{\partial s}
     \left( 
       s^3 h \langle{\rho}_\tx{eq}\rangle
       U_\tx{A}^2
       \frac{\partial}{\partial s} \frac{\langle \overline{u'_\phi} \rangle}{s}
     \right) 
 \equiv \frac{\partial}{\partial t} \frac{F_\tx{LR}}{s \langle{\rho}_\tx{eq}\rangle} \;,  \label{waveeq}
\end{equation}
where ${U_\tx{A}^2} = {\langle \overline{ \widetilde{B_s^2}} \rangle}/{\mu_0 \langle{\rho}_\tx{eq}\rangle}$,
implying a wave equation for angular velocity
 in the anelastic case \citep{JF15}.
\revtwo{We note that another possible definition would be $\langle \overline{\widetilde{B_s^2}}/\mu_0 \rho_\tx{eq} \rangle$ [see \ref{sec:momentum}],
 but this is less convenient in our formulation.}
\revoneR{As the restoring force of the wave is represented by $F_\tx{LR}$,
 the remaining terms} of the Lorentz force can be summed up to
 \revone{a term $F_\tx{LD} = F'_\tx{L} - F_\tx{LR}$.
This term} 
 represents the convection-driven fluctuations, 
 which interact with the magnetic field
 to drive the TWs through the Lorentz force \revone{and
 to modify their waveforms and/or speeds.
Below we see the latter effects but they are minor in our simulation,
 so we call $F_\tx{LD}$ a driving term}.
\revoneR{As we will see below,} the waves can also be driven by convective perturbations
 in the Reynolds force denoted by  $F'_\tx{R}$.  

A perturbation of angular velocity, ${\langle \overline{u'_\phi} \rangle}/{s}$,
can propagate in
 $s$ with Alfv\'{e}n speed, $U_\tx{A}$.
The speed depends on \revone{the magnitude of the background field,
 ${\widetilde{B_s^2}}$, and the density, ${\rho}_\tx{eq}$},
 both of which vary with $s$.
This special mode is nondispersive,
 i.e.~the speed is independent of wavenumbers.
Since the equation allows both inward and outward propagation,
 a superposition of those modes 
 could yield standing waves and enable normal mode solutions.
However, observational data for Earth, and numerical simulations,
 indicate a preference for (outwardly) propagating waves over standing ones
 (sec.~\ref{sec:intro}).

\section{Numerical simulations}
\label{sec:simulations}

\subsection{Model description} \label{sec:model}

\revoneR{To explore excitation of TWs in the gas giant we
 adopt Jovian dynamo models,}
%,
 which were built by \citet{J14} (hereafter
 referred to as J14) and \revoneR{developed by} \citet{DJ18}:\revoneR{see {J14} for the detailed description of the model set-up.}
The models explore the self-generation of magnetic fields by anelastic
 fluid motions in rotating spherical shells.
 The equilibrium reference state calculated by \citep{FBLNBWR12} was used,
 and viscous and diffusion terms are taken into account.
The reference state density, ${\rho}_\tx{eq}$,
 electrical conductivity, ${\sigma}_\tx{eq}$,
 and temperature, ${T}_\tx{eq}$, arise from a composition comprising of 
 a metallic hydrogen region above a rocky core, \revoneR{taken in this model as} 
 $r \ge r_\tx{c}\,\sim 6.45\,\times\,10^6$\,m $\sim 0.09\,R_\tx{J}$,
 and its continuous transition to a molecular hydrogen region.
The transition begins \revtwo{at $r \sim 0.85$-$0.90\,R_\tx{J}$}
 and only the region below a cut-off level,
 $r \le r_\tx{cut} \sim 6.70 \times 10^7$ m $\sim 0.96\,R_\tx{J}$, is
 modelled in our simulations,  \revoneR{the cut-off being required for numerical reasons}.
The density scale height, $N_\rho = \ln{[{\rho}_\tx{eq}(r_\tx{c})/{\rho}_\tx{eq}(r_\tx{cut})]}$,
 between the core boundary and the cut-off radius is approximately $3.08$.
Convection is largely driven by a uniform entropy source,
 which is released as the planet cools;
 this differs from the present geodynamo, which is primarily driven by
 buoyancy sources arising from the inner core boundary due to its
 freezing.

\revone{As the electrical conductivity, ${\sigma}_\tx{eq}$,
 drops by more than five orders from the metallic to the molecular region,} 
 a poorly-conducting layer is formed at the outermost part of the shell. 
\revone{Despite compressibility, the Proudman-Taylor constraint still strongly influences fluid motions in the outer layers when electrical conductivity is negligible.} 
%This 
\revoneR{The constraint is relaxed in the conducting region
 and this}
 produces a second imaginary cylinder, aligned with the rotation axis,
 that circumscribes the metallic hydrogen region which we call the magnetic tangent cylinder (MTC) \citep{DJ18},
 \revone{located at $s \sim 0.85$-$0.90\,R_\tx{J}$}.
This is in addition to the traditional 'kinematic' TC
 found at $s = r_\tx{c} \revtwo{= 0.0963\,r_\tx{cut}} \equiv s_\tx{tc}$, circumscribing the solid core. 
\revone{Unlike the kinematic TC, the MTC is not precisely defined, as the conductivity drop occurs over a finite radius range, 
but this range is thin enough for the MTC concept to be useful here:} \revoneR{for our purposes, we denote 
 $s_\tx{mtc}$ as the minimal $s$ at which the magnetic diffusion term becomes comparable 
 to the other terms in (\ref{waveeq}), which in our model is at  $\sim 0.89\,r_\tx{cut}$.}
The Jovian core leaves only a small fraction of the domain inside the TC.
We shall concentrate on the region outside the TC but inside the MTC,
 i.e.~$s_\tx{tc} \le s \le s_\tx{mtc}$.

We select three (models A, E, and I) out of nine models examined by {J14},
 which differ only in model parameters and \revone{entropy outer boundary conditions}. 
The chosen models and key quantities are listed in table \ref{table:dynamo_simulations}.
The global Rossby number, $Ro$, quantifying the relative strength of the inertia
 to the Coriolis force, is shown to be no greater than $5 \times 10^{-3}$.
The Elsasser number, $\Lambda$, \revoneR{is a dimensionless measure of the magnetic field strength} 
and is found to be
 approximately 6-10 in our simulations.
\revoneR{These yield Alfv\'{e}n numbers ranging from 0.45 to 0.62
 (see the caption of table \ref{table:dynamo_simulations})
 and the Alfv\'{e}n speed is faster than the rms velocity overall.}
Model I was reported to reproduce a magnetic field morphology broadly
 resembling that measured by Juno \citep{JH17}. 
Some smaller scale features of Jupiter's magnetic field as
 revealed by the mission more recently \citep{C18} differ from the models,
 notably in the equatorial asymmetry of the small scale field.
However, wave propagation is determined mainly by the large scale
 magnetic field,
 and TWs involve averaging over cylinders passing through both hemispheres. 
So this refinement of the Jovian magnetic field is not likely to affect our results greatly.

\begin{figure}[bh]
\includegraphics[width=105mm]{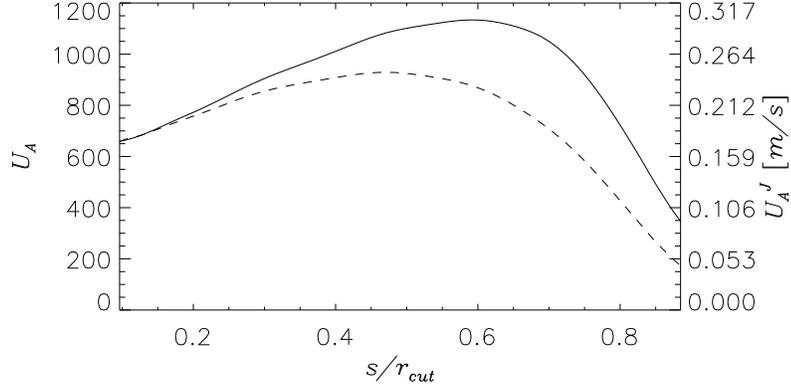}
 \caption{Alfv\'{e}n speed $U_\tx{A}$ (solid curve)
 as a function of cylindrical radius, $s$, for dynamo run I. 
 The abscissa is normalised by our cut-off radius, $r_\tx{cut} \sim 0.96\,R_\tx{J}$.
 \revtwo{The speed using a constant mid-radius value of the density
 is also plotted by a dashed curve for comparison.} 
 \revone{A MTC forms at $s \gtrsim 0.89\,r_\tx{cut} \sim 0.85\,R_\tx{J}$; 
 only the region outside the kinematic TC but inside the MTC
 ($0.096 \le {s}/{r_\tx{cut}} \le 0.89 \equiv s_\tx{mtc}/{r_\tx{cut}}$) is shown.} 
 The axis on the right-hand side indicates a dimensional scale
 $U_A^\tx{J}$ in metre per second:
 see the main text or table \ref{fig:axisymmetic_part}
 for the time unit used for the conversion.}
 \label{fig:TW_speeds}
\end{figure}

The magnetic fields self-generated in our simulations are
 \revone{non-reversing and dipolar during the simulations}.
\revone{They act as the background field for the MHD wave motions} discussed below.
\revone{The propagation speed of TWs is determined by the cylindrically averaged $B_s$ field.}
In figure \ref{fig:TW_speeds}, a solid curve depicts 
 the nondimensional Alfv\'{e}n speed,
 $U_\tx{A}$, as a function of cylindrical radius, $s$, 
 normalised by the cut-off radius, $r_\tx{cut}$, for model I.
Here the time and length are scaled by the magnetic diffusion time and
 the shell thickness ($D=r_\tx{cut} - r_\tx{c}$), respectively,
 and the bounds for $z$-averages are taken at $r_\tx{cut}$.
In the figure, the dashed line represents the Alfv\'{e}n speed with the density taken to be 
its constant mid-radius value \revoneR{$\rho_\tx{eq}(r_\tx{c}/2 +r_\tx{cut}/2)$} in the definition of $U_A$.
\revoneR{The anelastic Alfv\'{e}n speed has a peak at ${s}/{r_\tx{cut}} \sim 0.6$. At ${s}/{r_\tx{cut}} \lesssim 0.6$,
the density ${\rho}_\tx{eq}$ decreases with radius $r$ and 
 the $z$-mean Alfv\'{e}n speed increases with $s$ because  
${U_\tx{A}^2} = {\langle \overline{ \widetilde{B_s^2}} \rangle}/{\mu_0 \langle{\rho}_\tx{eq}\rangle}$.
At ${s}/{r_\tx{cut}} \gtrsim 0.6$, the density decrease effect is countered by the drop in
$\langle \overline{ \widetilde{B_s^2}} \rangle$ due to the field morphology, 
so for larger $s$, the speed gradually decreases  as the MTC is approached and crossed.}
\revoneR{Profiles of $U_A$ are similar for the other simulations explored here,
 with peaks at $0.6 \lesssim {s}/{r_\tx{cut}} \lesssim 0.7$.}
Table \ref{table:dynamo_simulations} also lists
 the speeds ${U_\tx{A}(s_\tx{mtc})}$ at the MTC radius
 and the expected traveltimes $\tau_\tx{A}$
 from the core boundary $s_\tx{tc}$ to the $s_\tx{mtc}$.
The speeds $U_\tx{A}$ are used for conversion to our dimensional time
 unit below \revoneR{(see details in sec.~\ref{sec:rescaling})}:
 a Jovian scale ${U_\tx{A}^\tx{J}}$ is shown on the right-hand side
 of the axis in fig.~\ref{fig:TW_speeds}.

\subsection{Internal dynamics: zonal flow fluctuations and their excitation} \label{sec:internal_region}

The time averaged components of azimuthal velocity,
 $\langle \overline{\widetilde{u_\phi}} \rangle$,
 show one very strong prograde jet outside the MTC and
 \revtwo{rather incoherent mean alternating flows within it}
 (figure 6 of {J14}).
In spite of the presence of generated magnetic fields and anelasticity,
 axisymmetric zonal flows inside the MTC still retain a significant fraction
 of the $z$-independent part of the flow.
By removing the mean part,
 we identify fluctuations of azimuthal flows,
 $\langle \overline{u'_\phi} \rangle$, which are of interest here.
Figure \ref{fig:axisymmetic_part} displays contours of
 $\langle \overline{u'_\phi} \rangle$ in $s$-$t$ space for the
 three runs.
In each diagram, white curves indicate the calculated Alfv\'{e}n speed,
 $U_\tx{A}$, to compare with the computed fluctuations.
A dimensional time is shown on the top of each image
 (details in the following section).
Run A (fig.~\ref{fig:axisymmetic_part}a) shows that some disturbances
 emerge near ${s}/{r_\tx{cut}} \gtrsim 0.6$
 and move outward to the poorly conducting layer;
 they can also be found to travel inwards towards the core boundary.
Their propagation speeds fit well with the predicted $U_\tx{A}$,
 suggesting that they are anelastic torsional Alfv\'{e}n waves.
\revone{They become more evident when filtered (see \ref{sec:axisymmetic_part_filtered}).}
Travelling TWs are found in Earth-like Boussinesq models
 \citep{WC10,TJT14,TJT15,SJNF17,TJT19}; 
 they mostly originate in the vicinity of the TC,
 where vigorous convection occurs near the solid inner core.
No obvious standing TWs have been found in geodynamo simulations to date.

Figure \ref{fig:axisymmetic_part}b displays contours of
 $\langle \overline{u'_\phi} \rangle$ for model E, where
 the relative strength of the viscosity to the Coriolis force is decreased.
We see significant fluctuations repeatedly occurring at an outer radius,
 $s/r_\tx{cut} \gtrsim 0.6$.
Interestingly, beneath the MTC,
 waves appear to form a node at around $s/r_\tx{cut} \sim 0.65$ and $\sim 0.8$, so 
figure \ref{fig:axisymmetic_part}b shows evidence of standing waves being excited in \revtwo{the Jovian models}.
We also find propagating features at a later time, $t \gtrsim 0.003$.
There are signatures of reflection, highlighted by the white lines,
 around the MTC at, for instance, $t \sim 0.0038$.

\revone{A simple one-dimensional model of Alfv\'{e}n waves propagating into a region
 where the diffusivity increases over a transition region was considered
 (not shown; see \ref{sec:appendix_1d} for the uniform diffusivity case). 
It shows that incident waves whose wavelength is shorter than,
 or comparable to, the thickness of the transition region are absorbed
 by diffusion, whereas waves with a wavelength longer
 than the transition thickness are mostly reflected. 
The theory also shows there is no phase change in
 $\langle \overline{u'_\phi} \rangle$, so a red patch in
 figure \ref{fig:axisymmetic_part}b should reflect into a red patch,
 as seen in the figure. 
When a wave packet reaches the MTC, the shorter wavelength components comparable
 to the (rather thin) transition region thickness are absorbed,
 while the longer wavelength components are reflected.}
This contrasts with the circumstances around Earth's CMB,
 which is a hard boundary of the fluid; 
 there a combination of the viscous dissipation and magnetic dissipation
 across the CMB controls the behaviour \citep{SJ16}.

In model I, \revone{where the entropy flux at $r_\tx{cut}$ is a given constant,}
 the nature of reflections from the MTC has been studied.
Figure~\ref{fig:axisymmetic_part}c shows
 the interaction of the blue feature with the MTC
 at $0.0008 \lesssim t \lesssim 0.0010$. 
\revone{Note that \revoneR{poor resolution of observational data
 and/or improper filters over them} may make the reflecting nature of the waves less clear (see \ref{sec:axisymmetic_part_filtered}).}
To describe the time evolution,
 we also present profiles of $\langle \overline{u'_\phi} \rangle$
 in figure \ref{fig:axisymmetric_waveforms}.
A trough \revoneR{came into existence} at $t\sim 0.0006$ and ${s}/{r_\tx{cut}} \sim 0.75$.
As time evolves, it eventually grows,
 while the waveform becomes sharper (fig.~\ref{fig:axisymmetric_waveforms}a).
\revone{This suggests a nonlinear influence on the TWs,
 arising from the terms $F_\tx{LD}$ and/or $F_\tx{R}$.} % \citep[e.g.][]{W74}.}
At $t \sim 0.00095$, 
 the trough reflects at around $s_\tx{mtc}$
 but also passes through the transition zone. 
The patterns of the incident and reflected waves 
 are compared in fig.~\ref{fig:axisymmetric_waveforms}b, which
 illustrates a positive reflection. 
However, there is a superposition of continuously excited waves, so 
 the amplitude and \revoneR{the} shape vary in our nonlinear simulation, and it is
 hard to determine the reflection coefficient or the phase shift accurately.
An abrupt change in a $U_\tx{A}$ profile may also yield reflections \cite[e.g.][]{AF63}.
Forward simulations of linear, nondissipative TWs in spherical geometry
 reported internal reflections where the gradient of a background
 magnetic field was steep \citep{CLM14}.
The role of the background velocity on the reflections in our simulations
 has not yet been elucidated.

The excitation mechanism of the waves is investigated in
 figures \ref{fig:excitation}a and b which display the forcing terms
 $F'_\tx{R}$ and $F_\tx{LD}$, respectively, for run I.
The Reynolds force is found to be important in the outer regions, $0.6\;r_\tx{cut} \lesssim s \lesssim s_\tx{mtc}$,
 whereas the Lorentz force is more evenly spread throughout the region and so more dominant
in the interior. 
This is understandable as the cylinders defining the wave motion
which have larger $s$ have a greater proportion of area in the vigorously convecting outer layers.
The term $F'_\tx{R}$ better matches the location and time
 at which $\langle \overline{{u'}_\phi} \rangle$ disturbances begin to travel than  $F_\tx{LD}$.
 This is in spite of the small \revoneR{global} Rossby number; 
 such an initiation was pointed out in Boussinesq cases \citep{TJT14}.
The convective motions are most vigorous in the outer layers of our Jupiter
 models, similar to fig.~6d of {J14};
 the density stratification enhances convective velocities to 
 get the heat flux out.
This produces a convergence of the Reynolds stress,
 particularly through the $\rho u_s u_\phi$ term,
 and continuously forces fluctuations, $\partial u_\phi/\partial t$,
 by almost-hydrodynamic Rossby waves\revoneR{, 
 which are found to be faster than the Alfv\'{e}n waves in the simulations by at least a factor 10 (not shown)}.
TWs in model A are also predominantly excited by $F'_\tx{R}$;
 it is rather mixed with Lorentz terms $F_\tx{LD}$ in model E.

Models for Earth, by comparison, have shown that the driving of TWs is possible 
 by either the Reynolds or Lorentz force. Geodynamo simulations have often shown the Reynolds 
 force to be the largest contributor \cite{TJT14}. However, as parameters are moved towards 
 their Earth-like values, geodynamo and magnetoconvection simulations \cite{TJT15} 
 display a growing influence of the Lorentz force. This is to be expected as the 
 role of the magnetic field increases as a balance closer to magnetostrophy is achieved. 
 The models for the Jovian dynamo discussed in this work use moderate values of the Ekman number, $E$, representing the ratio of the viscous force to the Coriolis force, 
 and the resulting Elsasser number is smaller than was possible in \cite{TJT15}.
It is not yet clear
whether TWs in Jupiter will be primarily driven by Reynolds or Lorentz force, and possibly both will be significant.
In Earth, Lorentz force will dominate, but in Jupiter convective velocities increase with radius. It is possible
strong convection in the outer regions could provide a significant contribution to the driving over the whole MTC, even though the density in these upper regions is small. 
Further simulations at lower $E$ and greater field strengths are necessary to decide this driving mechanism issue.

\section{\revone{Application to Jupiter}} \label{sec:detectability}

\subsection{\revone{Rescaling to the dimensional unit}} \label{sec:rescaling}

To examine whether signals due to TWs may be detectable in observational data, 
 we first convert the nondimensional time in our simulations
 to a dimensional unit.
\revoneR{Current numerical models
 \revone{are limited to \revoneR{numerically} accessible parameters \cite{SJNF17,A18},}
so parameters relating to diffusive processes have artificially increased values. 
No choice of dimensional units correctly scales all the physical processes
involved, so we choose scalings which aim to get the most important aspects of TWs right.} 
\revone{We choose the magnetic field scale
 by equating the field strength at the magnetic outer boundary in simulations
 to the observed outer boundary value \citep{TJT14}.}
Since the density is quite well-known, this gives the Alfv\'{e}n speed
 and hence a conversion between dimensionless and dimensional time.

Jovimagnetic models show the magnitude of the radial component 
 to be no greater than 60\,G, or 6\,mT,
 on a surface of $r \sim 0.85\,R_\tx{J}$: 
 in the equatorial region it is seemingly no greater than 30\,G and
 weaker than 1\,G for large regions \citep{C18}.
Taking 30\,G as a reasonable maximal field magnitude at our MTC radius,
 $s_\tx{mtc}$,
 and the density, ${\rho}_\tx{eq}$,
 of $8.53 \times 10^2$\,kg\,m${^{-3}}$, 
 then an Alfv\'{e}n speed, $U_\tx{A}^\tx{J}$, at this radius is approximately 
 $9.16 \times 10^{-2}$\,m\,s${^{-1}}$.
By matching this value with those of our simulations,
 our dimensional time unit $\tau_\tx{unit}$ is calculated
 through ${D U_\tx{A}}/{U_\tx{A}^\tx{J}}$,
 where $D$ is the shell thickness of $6.06 \times 10^7$\,m.
From this we calculate dimensional versions,
 ${\tau^\tx{J}}$ and ${\tau_\tx{A}^\tx{J}}$, of the analysed interval, $\tau$,
 and the TW traveltime, ${\tau_\tx{A}}$, respectively.
Values for each run
 are listed in table \ref{table:dimensional_outputs}.
While time units vary from 6.1 to 8.8 thousand years
 (as do analysed time windows ${\tau^\tx{J}}$ from 31 to 44 years),
 the traveltimes ${\tau^\tx{J}_\tx{A}}$ all fall within a 9-13 year window.

\revone
{A difficulty {for our scaling} arises
 when converting the averaged azimuthal velocities into dimensional units. 
The typical convective velocity at $0.85\,R_J$ is believed to be
 around $10^{-2}$\,m\,s${^{-1}}$ \citep{JH17}, but the surface equatorial
 zonal flow is nearly $100$\,m\,s${^{-1}}$. 
Simulations do get zonal flows that are larger than the convective flow,
 but they cannot yet reach the $10^4$ ratio due to the enhanced viscosity
 in the models, so it is uncertain how the axisymmetric azimuthal flow
 at depth should be scaled. 
Taking the unit of velocity as ${D}/{\tau_\tx{unit}}$ gives
 $3.15 \times 10{^{-4}}$\,m\,s${^{-1}}$ for run A, and this is the unit
 used for the averaged azimuthal flow in the figures. 
This gives a rather large convective velocity estimate
 of $0.5$\,m\,s${^{-1}}$ but a reasonable estimate of the mean zonal flow
 at \revoneR{$0.96\,R_J$} of about $2$\,m\,s${^{-1}}$
 (table \ref{table:dimensional_outputs}).
If we use a longer dimensionless time unit which puts the convective flow
 at $10^{-2}$\,m\,s${^{-1}}$, the amplitude of the azimuthal flow
 is reduced by a factor of around 50.} \revoneR{We prefer the shorter time unit,
 as we believe that future less diffusive models will have a higher ratio
 of zonal flow to convective flow,
 allowing a convective velocity of $10^{-2}$\,m\,s${^{-1}}$
 with a zonal flow of $\sim 2$\,m\,s${^{-1}}$ at $0.96\,R_J$.}

\subsection{\revone{Length-of-day variation \revoneR{(LOD)}}}  \label{sec:lod_variation}

Fluctuations in axisymmetric zonal flows produce variations in
 the angular momentum of the metallic hydrogen region
 which can be transferred to other parts of the planet.
This may produce fluctuations of the rotation period of the gas giant,
 namely LOD:
\makeatletter
\revoneR{\revone{this is often defined with the magnetic field (System \@Roman{3})
 that is generated in the metallic region. 
In Earth, in contrast,
 the LOD is fixed to the reference frame of the mantle.}}
\makeatother
\revtwo{Earth's LOD variation} with a period of nearly six years with  
 amplitude $\mathcal{O}(10^{-4}\,\textrm{s})$
 has been identified; its origin could be angular momentum exchange
 between the fluid core and the rocky mantle 
 through MHD waves (\citep{GJCF10}; sec.~\ref{sec:intro}).
One may envisage an analogous coupling in Jupiter
 between the deeper conducting metallic region and the overlying
 transition-molecular envelopes, as well as a Jovian LOD fluctuation.

We evaluate the influence by calculating the axial angular momentum change
 that is deduced from the axisymmetric disturbances in our metallic
 hydrogen region,
 \begin{equation}
  \delta \sigma = 2\pi \int_{s_\tx{tc}}^{s_\tx{mtc}}
   \int_{z_{-}}^{z_{+}} h \langle {\rho}_\tx{eq} \rangle s^2 \langle
   \overline{u'_\phi} \rangle\,dz\,ds \; ,\label{eq:delta_sigma}
 \end{equation}
 and those outside the region,
 \begin{equation}
  \delta \sigma_\tx{omtc} = 2\pi \int_{s_\tx{mtc}}^{r_\tx{cut}}
   \int_{z_{-}}^{z_{+}} h \langle {\rho}_\tx{eq} \rangle s^2 \langle
   \overline{u'_\phi} \rangle\,dz\,ds \; .
 \end{equation}
In figure \ref{fig:lod_fluctuations}
 the solid and dotted curves display
 the time evolutions of $\delta \sigma$ and $\delta \sigma_\tx{omtc}$ in
 model E, respectively.
The $\delta \sigma$ of the conducting region
 shows a quasi-periodic variation,
 corresponding to the flow oscillations (fig.~\ref{fig:axisymmetic_part}b).
The evolution is almost perfectly anti-correlated
 with the change $\delta \sigma_\tx{omtc}$ of the outermost transition zone, \revone{as it should be since total angular momentum is conserved.
Of interest is the coupling mechanism across the MTC.
\revoneR{Our simulations indicate both the magnetic and dynamic terms play a role (not shown);
 it is however uncertain how the coupling arises.}}
\revoneR{TWs with a short wavelength in the $s$-direction will be damped out by 
magnetic diffusion as soon as they leave the metallic hydrogen region,
but longer wavelength TWs are damped less rapidly
 as a single wavelength could extend right across the transition region
 (e.g. \ref{sec:appendix_1d}),
 allowing the TWs to be seen at the surface of the planet.}

Using our standard time unit and the density
 ($\revoneR{{{\rho}_\tx{eq}(r_\tx{c}/2 +r_\tx{cut}/2)}} = 2.56 \times 10^3 \,\textrm{kg}\,\textrm{m}^3$),
 we convert the dimensionless $\delta \sigma$ of maximum amplitude 38.7
 to its dimensional version, $\delta \sigma^\tx{J}$ with maximum amplitude $\sim 1.58 \times 10^{32}$\,N\,m\,s.
Assuming a value of $I=2.56 \times 10^{42}\,\textrm{kg}\,\textrm{m}^2$
 for the moment of the inertia \citep{NBHR12} %\citep{FBLNBWR12}
 and a daily period of $P= 3.57 \times 10^4$\,s for the planet,
 the change $\delta \sigma^\tx{J}$ is equivalent to 
 a period $\delta P$ of approximately 13\,ms.
Here $\delta \sigma^\tx{J} = I \delta \Omega = {-2\pi I \delta P}/{P^2}$, where
 $\Omega$ is the angular velocity.
If we use the longer dimensionless time unit which puts the convective flow at $10^{-2}$\,m\,s${^{-1}}$,
the amplitude of the period change is reduced by a factor of around 50 (sec.~\ref{sec:rescaling}). In 
table \ref{table:dimensional_outputs} we give this alternative scaling,
 which for model~E gives about 0.25\,ms, in brackets.

\makeatletter
Jupiter's mean LOD is determined to a precision more accurate than seconds 
 using the System \@Roman{3}.
\makeatother
These estimates largely rely on measurements of the decametric radio emission
 from the magnetosphere since the 1950s.
Decadal averages of the observed radio rotation period shows 
 its changes on the order of tens of milliseconds; 
 this remains the subject of some debate \citep{HCR96,RYK01,RH16}.
The measurements may reflect a time-varying SV due to unsteady convective flow,
 rather than changes being solely due to TWs. 
Our estimates indicate that TWs could be a part of LOD fluctuations,
 but separating the convective flow-induced changes from the TW changes
 will not be easy.

\subsection{\revone{Flow change above the metallic region}} \label{sec:surface_flow}

Unlike terrestrial planets, 
\revoneR{gas giants may allow deep-origin perturbations
to be observed at the surface.}
Figures \ref{fig:surface_flow_fluctuations}a and b show contours of 
 the fluctuating zonal flow $\overline{u'_\phi}$ 
 on the cut-off surface, $r_\tx{cut}$, in the northern and southern
 hemisphere for model E, respectively.
The latitude-dependent data is displayed in $s$-$t$ space
 to enable comparison with the figures and wave speeds shown earlier.
%Note that the plots include the region outside the MTC.
The amplitude is scaled by the maximum of the mean zonal flow
 $\overline{\widetilde{u_\phi}}$ on the same surface,
 which is the maximal speed of the prograde, equatorial jet reproduced
 in the simulation (sec.~\ref{sec:internal_region}).
Figs.~\ref{fig:surface_flow_fluctuations}c-d
 show the same data filtered to remove 
 all periods outside the range from $t = 0.00063$ to $0.0025$,
 i.e. from 5.6 years to 22 years in the dimensional unit for 30\,G.

In both hemispheres we find corresponding fluctuations on the surface,
 although they look much noisier than the internal wave motions.
The filter used in figures c-d helps to visualise the wave signals
 more clearly.
The variations are found to be almost symmetric with respect to the equator,
 which is a consequence of the predominantly $z$-independent flow.
Oscillations and both equatorward and poleward propagation
 are seen at mid and high latitudes where $s \lesssim s_\tx{mtc}$,
 whereas the equatorial region $s \gtrsim s_\tx{mtc}$
 features only equatorward migration. 
We interpret this as \revone{partial transmission}
 through the MTC and absorption within the resistive, transition layer
 (sec.~\ref{sec:internal_region}; \revone{\ref{sec:appendix_1d}}).
The abrupt change in zonal flow fluctuations on spherical surfaces
 signifies the location of the MTC. 
Thus it can act as an indicator of the location
 where the transition begins, i.e. the magnetic dissipation becomes significant.
This is however hard to identify when searching in a few snapshots only,
 as examined for zonal wind profiles;
 exploration in $\theta$-$t$ space - sometimes called
 a Hovm\"{o}ller diagram - is essential for the identification.

\revoneR{On the surface of our cutoff level}
 the maximum of the fluctuating velocity \revoneR{in model E}
 is 11\,\% of the mean velocity.
Other models show analogous fractions of 12-15\,\%;
 the values are listed in table \ref{table:dynamo_simulations}.
Converting to dimensional units as before,
 the fluctuation amplitudes at $r_\tx{cut} \sim 0.96\,R_\tx{J}$
 are found to be about 0.1-0.2\,m\,s${^{-1}}$ and 3-5 $\times 10^{-3}$\,m\,s${^{-1}}$
 for the equatorial field of 30\,G and 0.6\,G, respectively,
 whereas the mean velocities are an order greater
 (table \ref{table:dimensional_outputs}).

\revoneR{The surface zonal flows may extend deep into the interior \citep[e.g.][]{B76,JK09,HGW15}.
Jupiter's gravitational harmonics have recently been  obtained
 by the Juno mission, \citet{KGHetal18} and \citet{GMMetal18},
providing evidence that the zonal flows do go down to $\sim$ 0.95-0.97\,$R_\tx{J}$. 
To date, observational constraints on the speed of deep zonal flows are weak: \cite{GMMetal18} 
show that the zonal flow in the deep
interior must be less than 10\% of the cloud-level value, while \cite{KGHetal18} suggest that the
zonal flows could fall off exponentially with an e-folding depth of 1000-3000\,km to be consistent with the Juno gravity data.}
This is compatible with our higher velocity conversion range of $\sim 1.5$\,m\,s${^{-1}}$
 (see ${\overline{\widetilde{u_\phi}}}$ in table \ref{table:dimensional_outputs}).
At lower depths
 flow models inverted from the jovimagnetic SV
 suggest a velocity of order $10^{-2}$\,m\,s${^{-1}}$
 at the top of the expected conducting region (given $0.85\,R_\tx{J}$),
 not far from those estimated with scaling properties based on
 available heat fluxes \citep{RH16,YGCD13}.
Those deep dynamics may set the thermodynamical conditions
 at the cloud deck and trigger visible photochemical changes.

\revone{
Earth-based campaigns have monitored the long term variability
 of the cloud and/or atmospheric appearance of the gas planet.
Global upheavals are recurrent activities 
 spreading over several latitudinal bands  
 and occur at intervals of decades, irregularly in most cases \citep{R95,F17}.
At some epochs,
 5- or 10- years periodicities in jetstream outbreaks or
 fades/revivals were recognised at the North Temperate Belts (NTB), 23-35$^\circ$ N.
Recent datasets - primarily collected with
 the Hubble Space Telescope between 2009-2016 - 
 have been updated to identify the most relevant change in zonal winds
 near 24$^\circ$ N of about 10\,m\,s${^{-1}}$
 and 5-7 year periods at a few lower-latitudes \citep{TWPSetal17}.
These seem interesting in comparison to the internal flows simulated earlier. 
First, the amplitude of such disturbances is $\sim$ 10\,\%, or less,
 of the 150\,m\,s${^{-1}}$ stable jet \citep{TWPSetal17}.
Second, the NTB latitudes correspond to cylindrical radii
 $s \sim$ 0.82-0.92\,$R_\tx{J}$,
 which likely lie in the outermost part of the metallic region
 and the transition zone.
At these
 radii/latitudes, TWs in our models exhibited
 oscillations with periods of several years or longer and
 sometimes at irregular intervals.}

\section{Concluding remarks and discussion}
\label{sec:summary}

We have demonstrated, through our anelastic models,
 that torsional Alfv\'{e}n waves could
 be excited in Jupiter's metallic hydrogen region.
The axisymmetric MHD disturbances can propagate in \revone{cylindrical radius}
 on timescales of Alfv\'{e}n speeds \revone{in a medium with a variable
 equilibrium density}. 
In the Jovian dynamo models we adopted, waves were excited at the outermost part of
 the conducting region, where nonaxisymmetric convective motions were
 vigorous and the resulting stresses drove the axisymmetric fluctuations.
TWs were found to travel both
 outwards and inwards.
Modes propagating outwards were found to be partially transmitted to the poorly
 conducting layer but could also be reflected around the MTC. 
This results in waves travelling inwards, back into the deeper interior of the conducting region.
Since convection perturbs the fluid at all times 
 it is able to continuously supply a source for TWs travelling in both directions.
If their amplitude and timing matches, 
 a superposition of the opposed propagation enables the formation of standing waves,
 as observed in our model E.
Our simulations suggest there may be a 
mixture of travelling, reflecting, and standing waves in giant planets.
\revoneR{Our results suggest TWs in giant planets behave rather differently
from those in the Earth's core \cite{WC10,TJT14,TJT15,SJNF17,TJT19}. 
In geodynamo models the waves appear to be preferably excited at a location 
with vigorous convection near the inner core and they do not reflect upon impact with the CMB.}

A key requirement for reflection here is the existence of the MTC,
 which is created by the drastic decrease of the electrical conductivity in the gas giant.
The MTC may act as an interface for the waves approaching the magnetically-dissipative fluid layer, 
 which enables \revone{reflection as well as transmission}.
The interface created by the varying conductivity
 may allow reflection of waves to be a feature within Jupiter.
Whilst the size of the dynamo region is currently hard to define,
 detecting reflections from data may enable us to infer the radius where the transition from metallic to molecular hydrogen indeed begins.
This is analogous to how seismology has constrained the structure of the
deep Earth.

TW traveltimes across the metallic region were estimated at several years,
 provided that the time units in our simulations were chosen
 so that the Alfv\'{e}n speed at the equator at our MTC level
 matches that suggested by jovimagnetic models and adiabat
 density models.
An equatorial radial field of maximal strength 30\,G yields
 traveltimes of 9-13 years;
 longer timescales are feasible when a weaker field is implemented.

\revone{
The fluctuations of zonal flows yielded
 an exchange of angular momentum between the metallic
 hydrogen region and the overlying molecular regions. 
With the time units adopted,
 the waves could give rise to variations in the LOD
 no greater than $10^{-2}$\,s.
\revoneR{Alterations in Jupiter's radio rotation period might partly be due to true LOD changes as well as the magnetic SV. 
 We also note that uncertainties in the chosen scaling,
 which arose from the limitation of the current numerical dynamo models,
could affect our LOD variation estimates.}}

\revone{
Our simulations also demonstrated
 the wave motions identified through zonal flows
 on a spherical surface above the metallic region.
The surface fluctuations were sizable, up to 15\,\% of
 the maximal amplitude of the steady zonal component at $r \sim 0.96 R_J$.
The reflecting and/or transmitting nature across the MTC could be projected
 upwards from the metallic region to the surface.
Juno's gravity measurements have constrained interior models
 by identifying that visible surface zonal flows penetrate downwards significantly \citep{KGHetal18,GMMetal18}.
Assuming this deep origin, 
 variations at the cloud deck could display some evidence of TWs. }

%
%{
Another possible way to detect TWs is from the jovimagnetic SV,
 which is inferred at the top of the metallic region.
The projection from internal wave motions to the SV
 is rather complicated, as discussed in the context of Earth's fluid core.
TWs may contribute to the occurrence of geomagnetic jerks;
 they cannot however account for all phenomena alone (see \citep{MHPPJV10}
 for a review).
%To link wave motions to the magnetic SV, 
% zonal flow fluctuations seen at the surface are required to linearly
% induce rapid variations in the surface magnetic field.
%Here caution is necessary since such magnetic changes can be initiated by
% other terms in the induction equation.
Nevertheless, an increase of spatial and temporal coverage in magnetic data
 is expected to better resolve the SV and inverted flow models
 on the top of the metallic region.
The ongoing Juno magnetic measurements, coupled with theoretical studies,
 will offer a promising route to develop our knowledge
 on the dynamics in the dynamo region.
%}

\section*{Acknowledgments}

We acknowledge support from the Japan Society for the Promotion of
 Science (JSPS) under Research Activity Start-up No.~17H06859,
 as well as from the Science and Technology Facilities Council of the UK,
 STFC grant ST/N000765/1.
This work was undertaken on ARC2, part of the High Performance Computing facilities at the University of Leeds, UK. Also this work used the DiRAC@Durham facility managed by the Institute for Computational Cosmology on behalf of the STFC DiRAC HPC Facility (www.dirac.ac.uk).
This work was partly made during the visit at the Lorentz Center Leiden.
We also thank Amy Simon for helpful discussions.
Comments by Thomas Gastine and
 Johannes Wicht helped to improve the manuscript.

%\section{Bibliography styles}
%
%There are various bibliography styles available. You can select the style of your choice in the preamble of this document. These styles are Elsevier styles based on standard styles like Harvard and Vancouver. Please use Bib\TeX\ to generate your bibliography and include DOIs whenever available.
%
%Here are two sample references: \cite{Feynman1963118,Dirac1953888}.

\section*{References}

%\bibliography{mybibfile}

%%%%%%%%%%%%%%%%%%%%%%%%%%%%%%%%%%%%%
%%%%%%%%%%%%%%%%%%%%%%%%%%% tables
\clearpage
\begin{table}[th]
 \caption{Dimensionless input and output parameters of dynamo simulations selected
 from {J14} \citep{J14}.
The columns $E$, $Ra$, and $H$ list the Ekman number, the Rayleigh
 number, and the volumetric entropy source, respectively.
The kinetic and magnetic Prandtl numbers are fixed at $Pr=0.1$ and $Pm=3$,
 respectively.
\revone{Constant entropy on both boundaries, $r_\tx{c}$ and $r_\tx{cut}$, for
 models A and E;
 for model I, constant entropy on the inner boundary and fixed entropy-flux outer boundary.
In all three models,
 a stress-free outer boundary, a no-slip inner boundary, and electrically
 insulating conditions both at both boundaries are used.}
The columns $\tau$, $Ro$, and $\Lambda$ list the analysed time interval,
 the Rossby number (equal to ${u_\tx{rms} E}/{Pm}$ in our scaling
 with $u_\tx{rms}$ being the rms flow vigour), and the Elsasser number
 (equal to the rms field strength, $B_\tx{rms}$, in our scaling), respectively,
 \revtwo{where averages over the whole volume are taken.}
\revone{These are used to yield the Lehnert number
 $Le = {B_\tx{rms}}/{(\sqrt{\rho_\tx{eq} \mu_0}D \Omega)} = {\sqrt{{\Lambda E}/{Pm}}}$ of 6.6-7.3$\times 10^{-3}$
 and the Alfv\'{e}n number $A = {u_\tx{rms}\sqrt{\rho_\tx{eq}\mu_0}}/{B_\tx{rms}} = {{Ro}/{Le}}$ of 0.45-0.62.}
The columns ${U_A (s_\tx{mtc})}$, $\tau_\tx{A}$,
 $\delta \sigma$, ${\overline{u'_\phi}(r_\tx{cut})}$,
 and ${{\overline{u'_\phi}}/{\overline{\widetilde{u_\phi}}} (r_\tx{cut})}$
 represent, respectively, 
 the Alfv\'{e}n speed at the MTC,
 the traveltime across the conducting region
 (from the core boundary to the MTC),
 the maximal amplitude of the axial angular momentum fluctuation in the
 metallic region,
 the maximal fluctuating zonal velocity,
 and the fraction of the maximal fluctuation part 
 to the maximal time-averaged part at the cut-off boundary.}
\label{table:dynamo_simulations}
\centering\footnotesize
\vspace{3mm}
\begin{tabular}{lrrrr rrlllll}
\hline
 Run  & $E$ & $Ra$ & $H$ & $\tau$ 
      & $Ro$ & $\Lambda$ & $U_\tx{A}(s_\tx{mtc})$ & $\tau_\tx{A}$ & $\delta \sigma$ & ${\overline{u'_\phi}(r_\tx{cut})}$ & ${{\overline{u'_\phi}}/{\overline{\widetilde{u_\phi}}} (r_\tx{cut})}$ \\
\hline
 A & $2.5 \times 10^{-5}$ & $1.1 \times 10^7$  & 1.5 & 0.005
          & 0.0037 & 5.5 & 291 & 0.0016 & 37.4 & 772 & 0.145 \\
 E & $1.5 \times 10^{-5}$ & $2.0 \times 10^7$  & 1.4 & 0.005
          & 0.0041 & 8.8 & 421  & 0.00121 & 38.7 & 662 & 0.106 \\
 I & $1.5 \times 10^{-5}$ & $2.0 \times 10^7$  & 1.4 & 0.005
          & 0.0035 & 10.3 & 347 & 0.00120 & 42.2 & 686 & 0.115 \\
\hline
\end{tabular}
%\tablenotetext{a}{Footnote text here.}
\end{table}

\clearpage
\begin{table}[th]
 \caption{Dimensional parameters of the simulations.
 The first column ${\tau_\tx{unit}}$ lists the time unit
 which is represented in years
 and is used to convert our dimensionless time to its dimensional version. The
 assumption is made of an Alfv\'{e}n speed of
 $9.16 \times 10^{-2}$\,m\,s${^{-1}}$,
 or, equivalently, a radial field strength of 30\,G,
 at the equator at the top of the metallic region $\sim 0.85\,R_\tx{J}$;
 calculated as described in the main text. 
 The columns ${\tau^\tx{J}}$, ${\tau^\tx{J}_\tx{A}}$, ${\delta \sigma^\tx{J}}$,
 $\delta P$, ${\overline{u'_\phi}(0.96\,R_\tx{J})}$, and
 ${\overline{\widetilde{u_\phi}}(0.96\,R_\tx{J})}$ present
 the dimensional version of the analysed interval $\tau$,
 traveltime $\tau_\tx{A}$, maximal angular momentum $\delta \sigma$,
 maximal LOD variation, maximal zonal flow fluctuation
 $\overline{u'_\phi}$ at the surface $r_\tx{cut} \sim 0.96\,R_\tx{J}$,  
 and maximal mean zonal flow ${\overline{\widetilde{u_\phi}}}$
 at the same surface, respectively.
Figures in brackets ( ) are based on the longer time unit mentioned in the text: an Alfv\'{e}n speed of $1.83 \times 10^{-3}$\,m\,s${^{-1}}$, or a field strength of 0.6\,G, at $r \sim 0.85\,R_\tx{J}$ is assumed.}
\label{table:dimensional_outputs}
\vspace{3mm} 
\centering\scriptsize
\begin{tabular}{l rrr ll ll}
\hline
 Run  & $\tau_\tx{unit}$\,[$10^3$\,yrs] & ${\tau^\tx{J}}$\,[yrs] & ${\tau^\tx{J}_\tx{A}}$\,[yrs] &
 $\delta \sigma^\tx{J}$\,[$10^{32}$\,N\,m\,s] & ${\delta P}$\,[$10^{-3}$\,s] &
 ${\overline{u'_\phi} (0.96\,R_\tx{J})}$\,[m\,s${^{-1}}$] &
 ${\widetilde{\overline{u_\phi}}(0.96\,R_\tx{J})}$\,[m\,s${^{-1}}$] \\
\hline
 A  & 6.10  & 30.5  &  9.7 & 2.20 (0.044) & 18 (0.35) & 0.24 (0.0049) & 1.7 (0.033)\\
 E  & 8.81  & 44.1  & 10.7 & 1.58 (0.032) & 13 (0.25) & 0.14 (0.0029) & 1.4 (0.027)\\
 I  & 7.26  & 36.3  & 13.3 & 2.09 (0.042) & 17 (0.33) & 0.18 (0.0036) & 1.6 (0.031)\\
\hline
\end{tabular}
\end{table}

%%%%%%%%%%%%%%%%%%%%%%%%%%% figures 
%%%%%%%%%%%%%%%%%%%%%%%%
\clearpage

\begin{figure}
\revone{(a)}\\
\includegraphics[width=95mm]{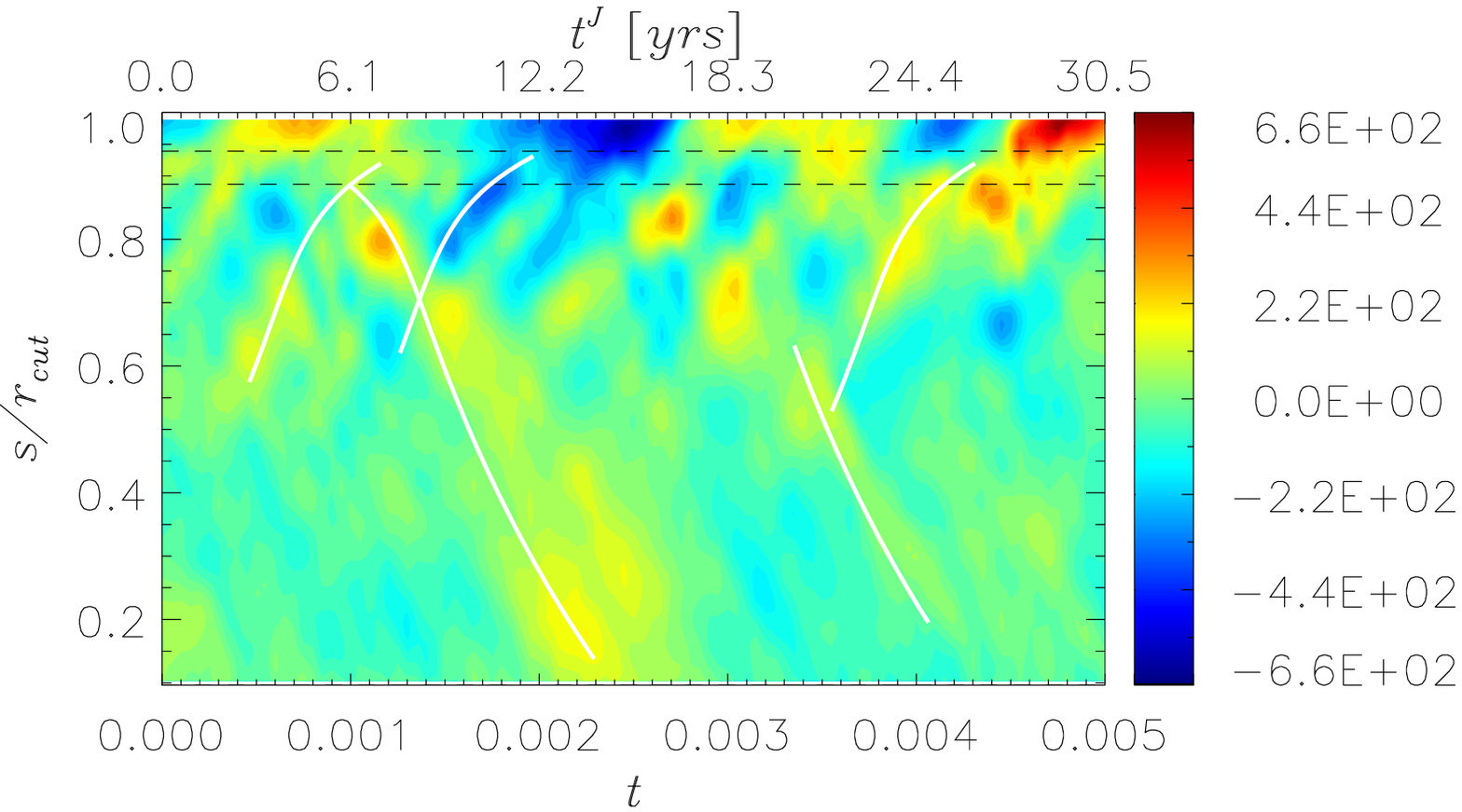} \\
\revone{(b)}\\
\includegraphics[width=95mm]{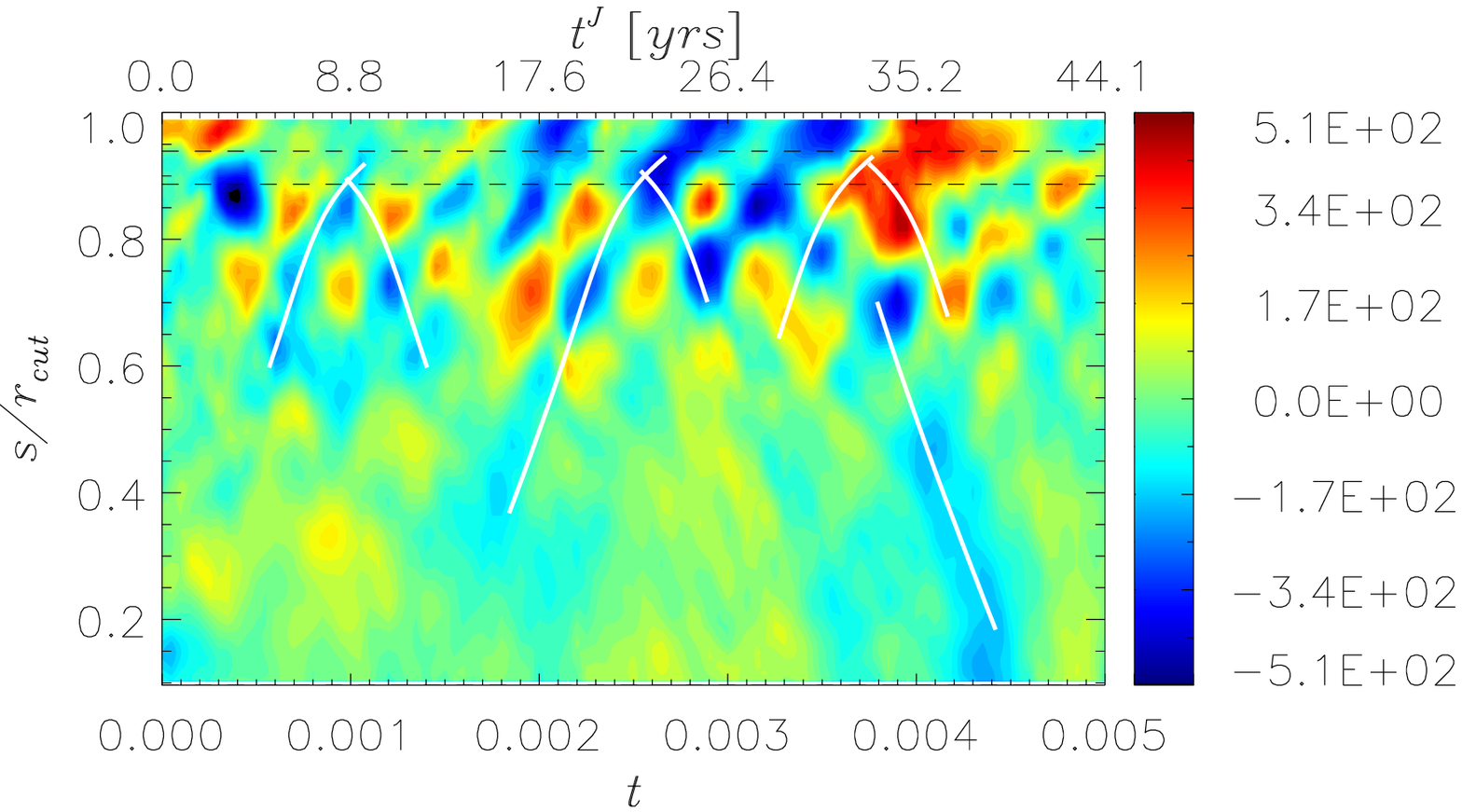} \\
\revone{(c)}\\
\includegraphics[width=95mm]{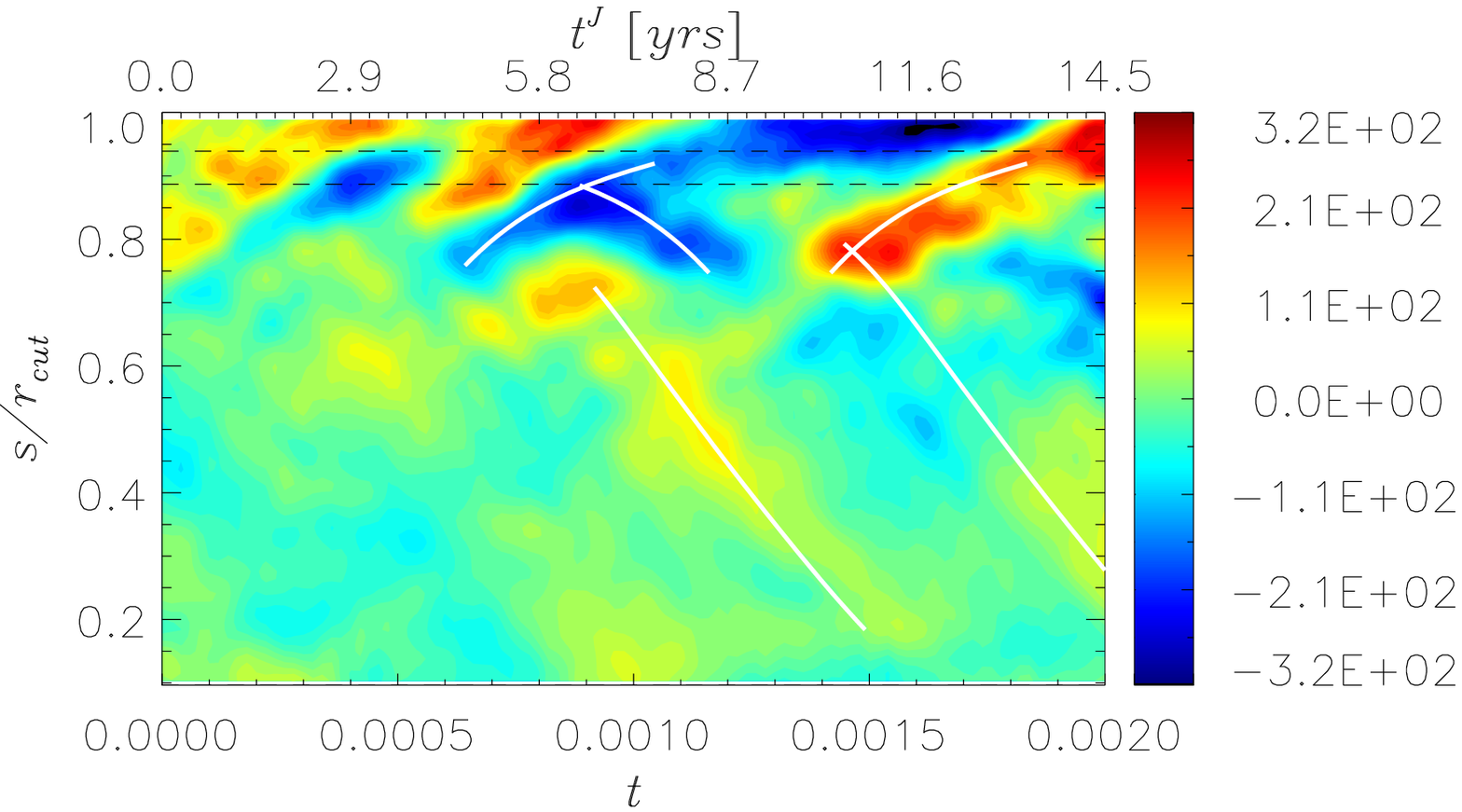}
 \vspace{-2mm}
 \caption{The fluctuating, $z$-averaged azimuthal velocity,
 $\langle \overline{u'_\phi} \rangle$, for run A (a), run E (b), and run I (c).
 White curves indicate
 phase paths of the Alfv\'{e}n speed, $U_\tx{A}$.
 A dimensional time scale, $t^\tx{J}$, is represented in years
 on the top of each panel.
 \revone{The horizontal dashed lines indicate a range of the MTC radius,
 ${s}/{r_\tx{cut}} \sim 0.89$ and $0.94$.}}
 \label{fig:axisymmetic_part}
\end{figure}

\begin{figure}
\revone{(a)}\\
\includegraphics[width=97mm]{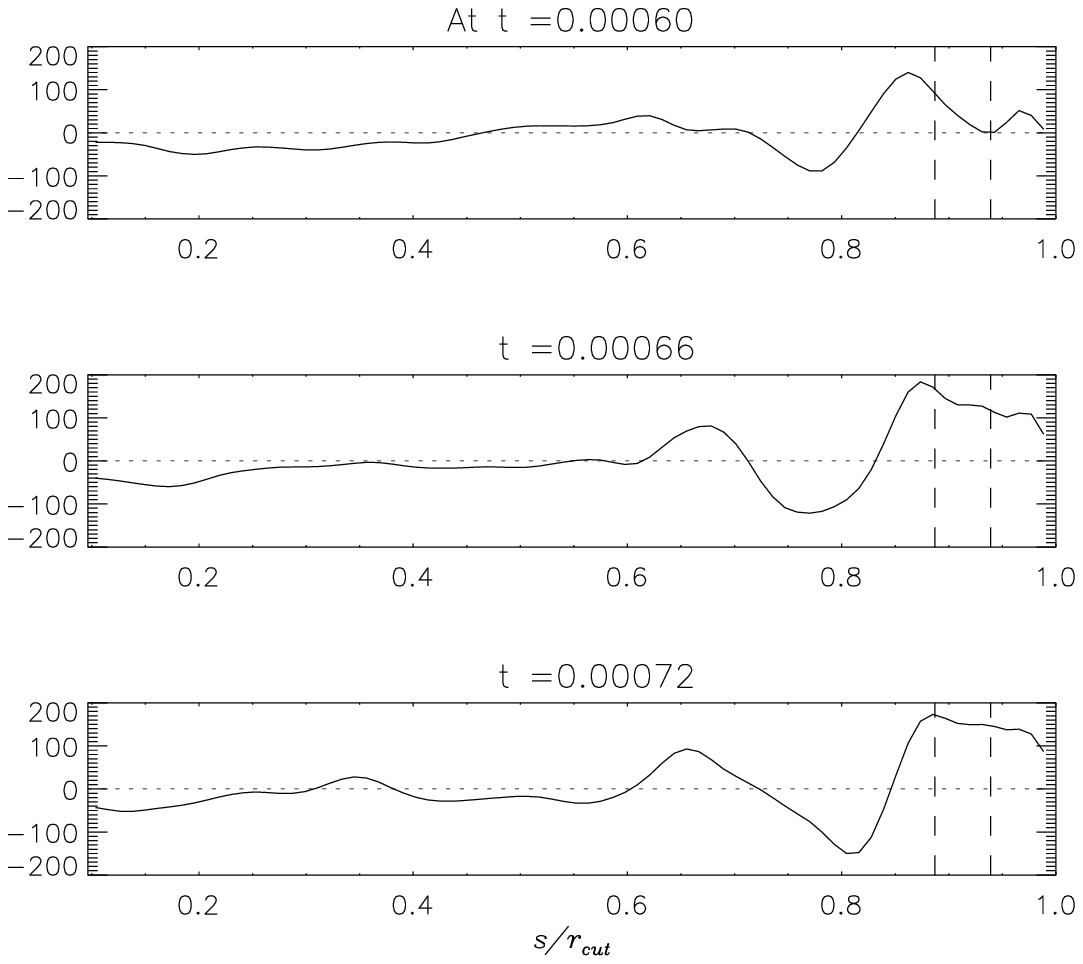}\\
\revone{(b)}\\
\includegraphics[width=97mm]{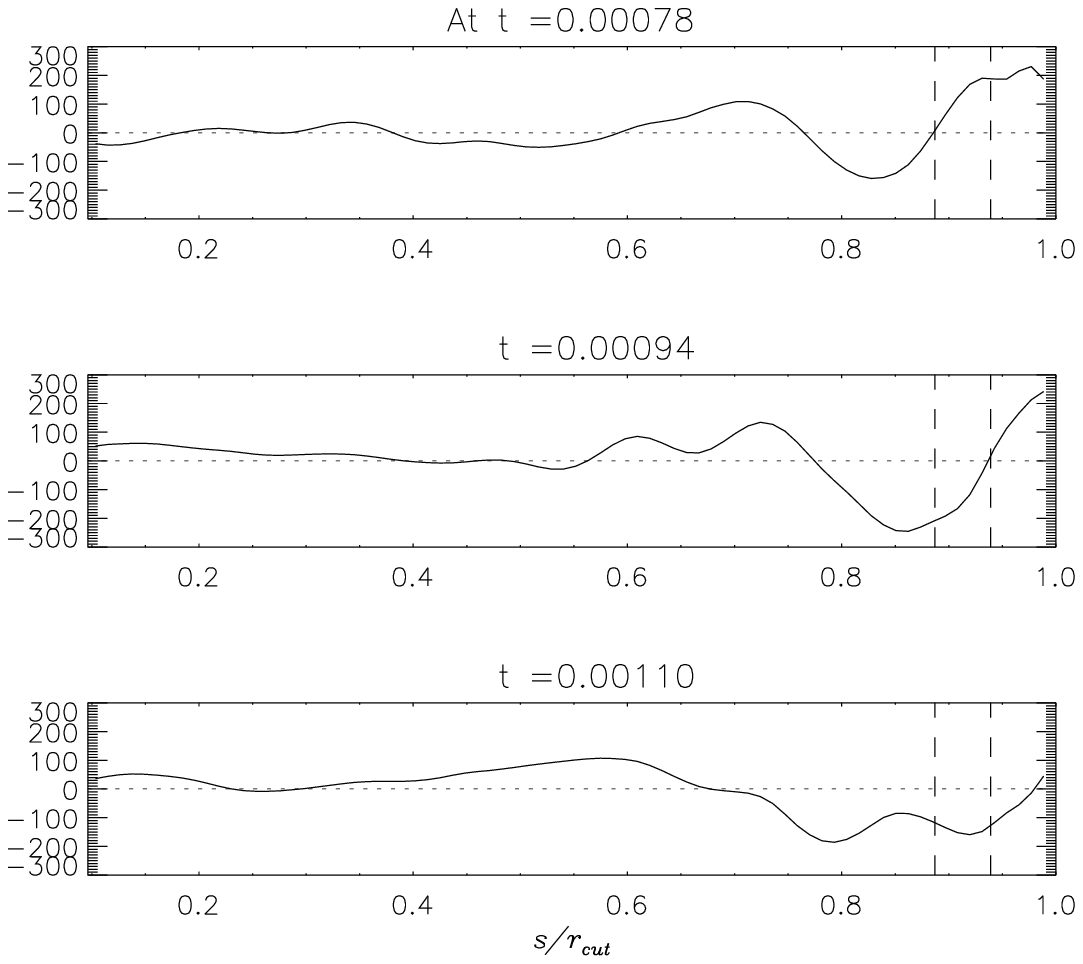}
 \caption{Time evolution of $\langle \overline{u'_\phi} \rangle$
 for run I displaying
 \revone{(a)} a progressing, steepening wave
 and \revone{(b)} waveforms before and after a reflection from the MTC.
 \revone{A range of the MTC cylindrical radii,
 ${s}/{r_\tx{cut}} \sim 0.89$ and $0.94$,
 are indicated by the vertical dashed lines in each panel.}}
 \label{fig:axisymmetric_waveforms}
\end{figure}

\clearpage
\begin{figure}
 (a)\\
 \includegraphics[width=95mm]{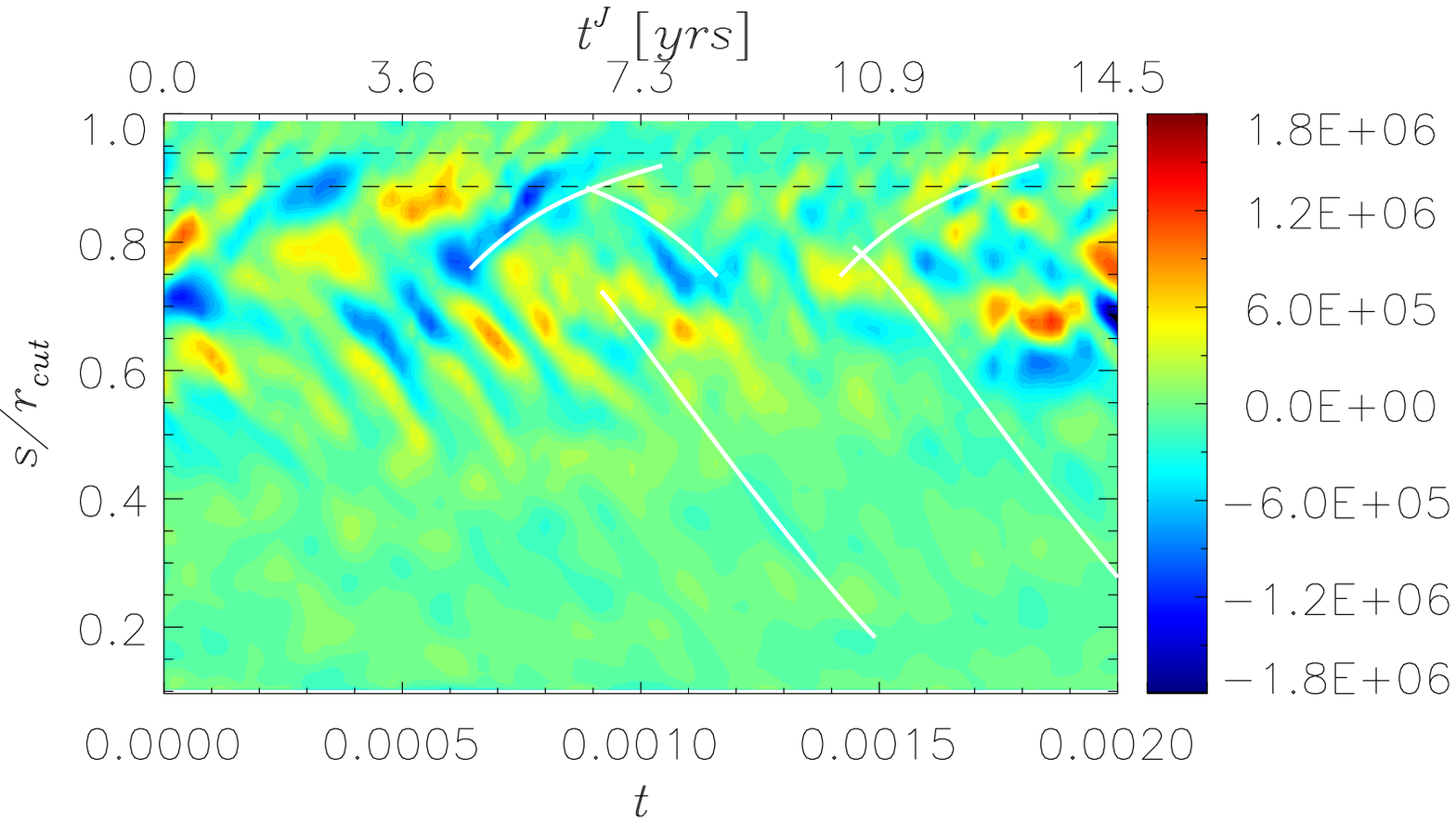}\\
 (b)\\
 \includegraphics[width=95mm]{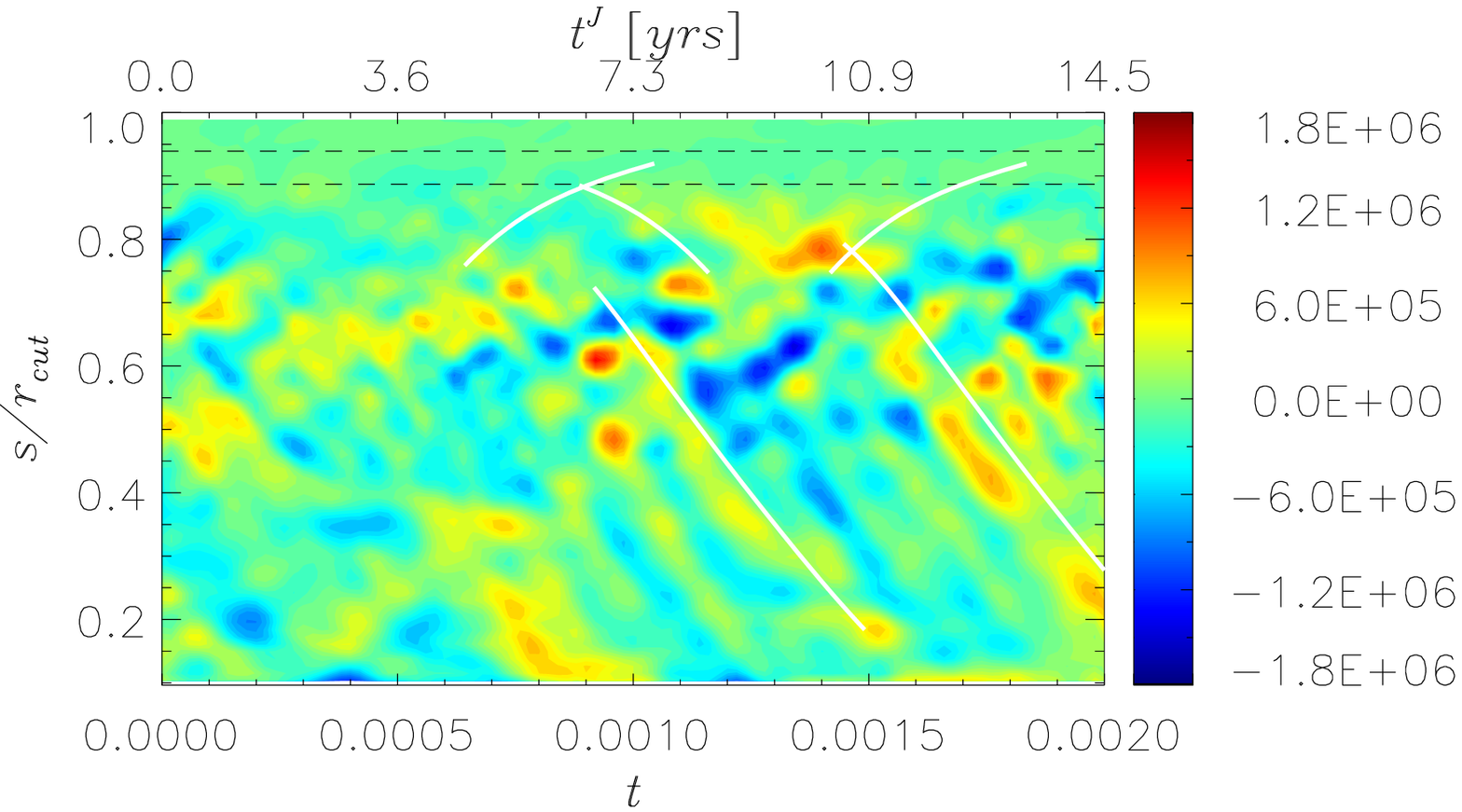}
 \caption{Forcing terms, (a) $F'_\tx{R}$ and (b) $F_\tx{LD}$,
 in the azimuthal momentum equation for run I.
 White curves represent the same phase paths as those shown
 in fig.~\ref{fig:axisymmetic_part}c.
 \revone{The horizontal dashed lines indicate the range of the MTC radii.}}
 \label{fig:excitation}
\end{figure}

\begin{figure}
 \includegraphics[width=100mm]{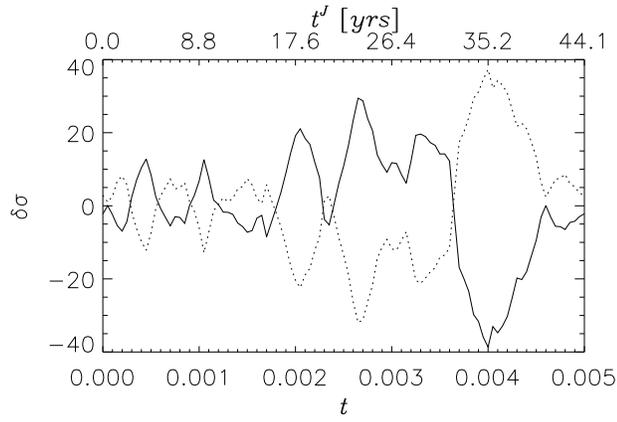}
 \caption{The axial angular momentum change for run E. 
 The solid curve represents $\delta \sigma$ (for the metallic region;
 ${s_\tx{tc}} \le s \le {s_\tx{mtc}}$),
 whilst the dotted curve shows $\delta {\sigma_\tx{omtc}}$ (for
 the transition zone; ${s_\tx{mtc}} < s \le {r_\tx{cut}}$).
 Their anti-correlation shows the angular momentum exchange between the
 regions. \revone{They are locally peaked at $t \sim 0.00045, 0.00105, 0.00205, 0.00265, 0.0033,$ and $0.0046$: at the respective times the internal waves in fig.\ref{fig:axisymmetic_part}b have crests (troughs) at $0.7 \lesssim {s}/{r_\tx{cut}} \lesssim 0.8$ ($0.8 \lesssim {s}/{r_\tx{cut}} \lesssim {s_\tx{mtc}}$).}}
 \label{fig:lod_fluctuations}
\end{figure}

\begin{figure}
 \begin{tabular}{ll}
 (a) & (c) \\
 \includegraphics[width=75mm]{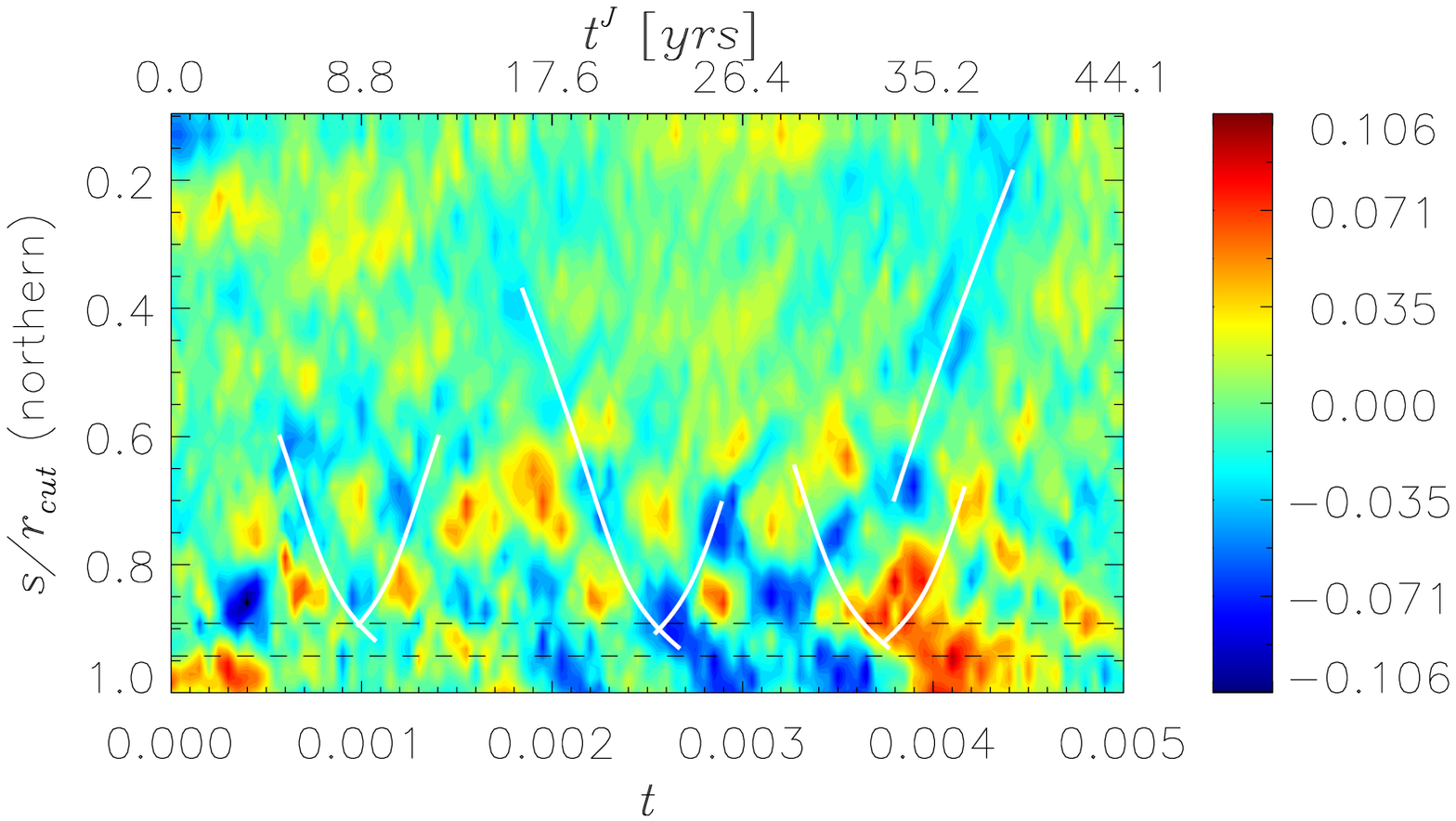}&
 \includegraphics[width=75mm]{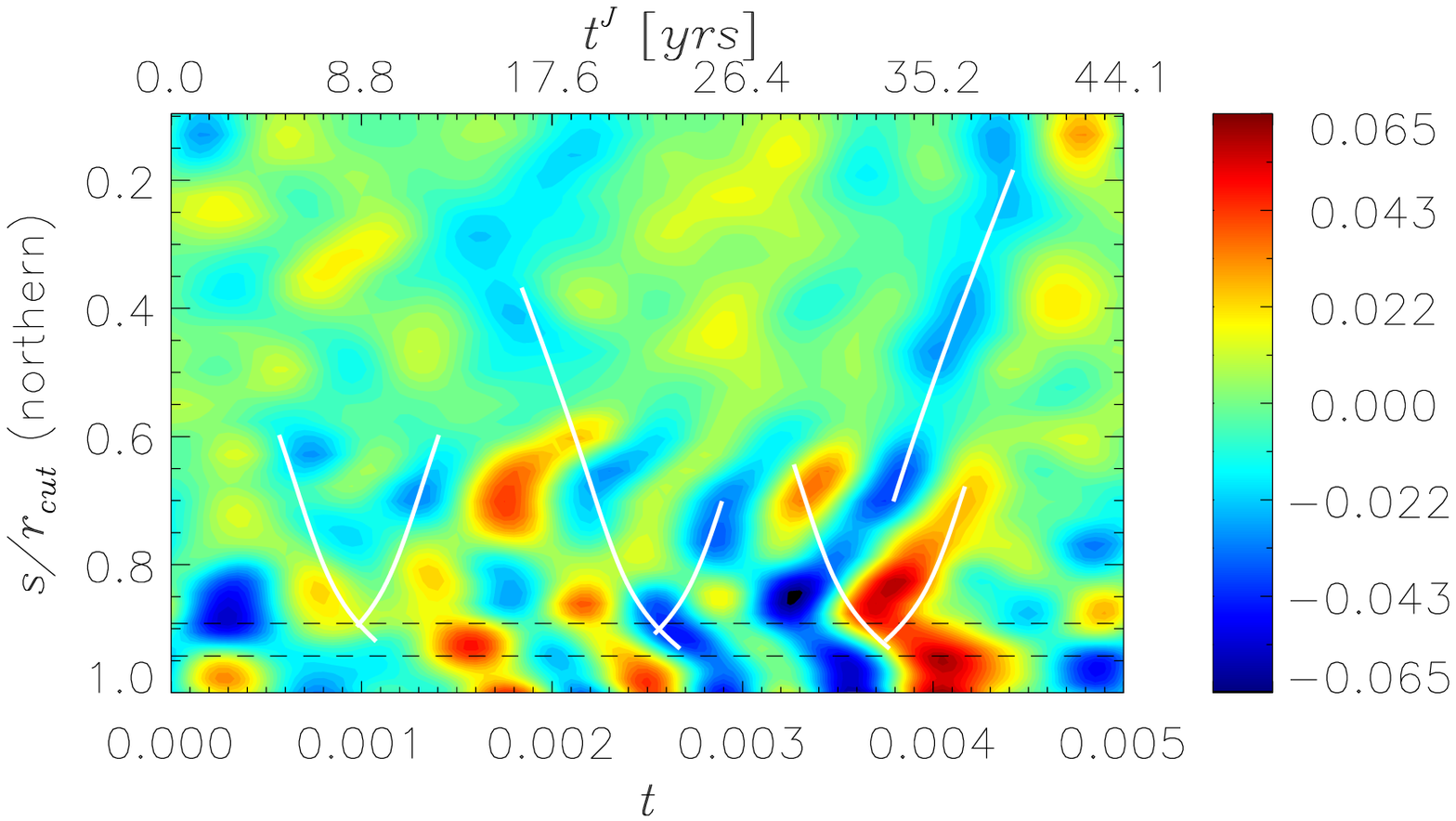}\\
 (b) & (d) \\
 \includegraphics[width=75mm]{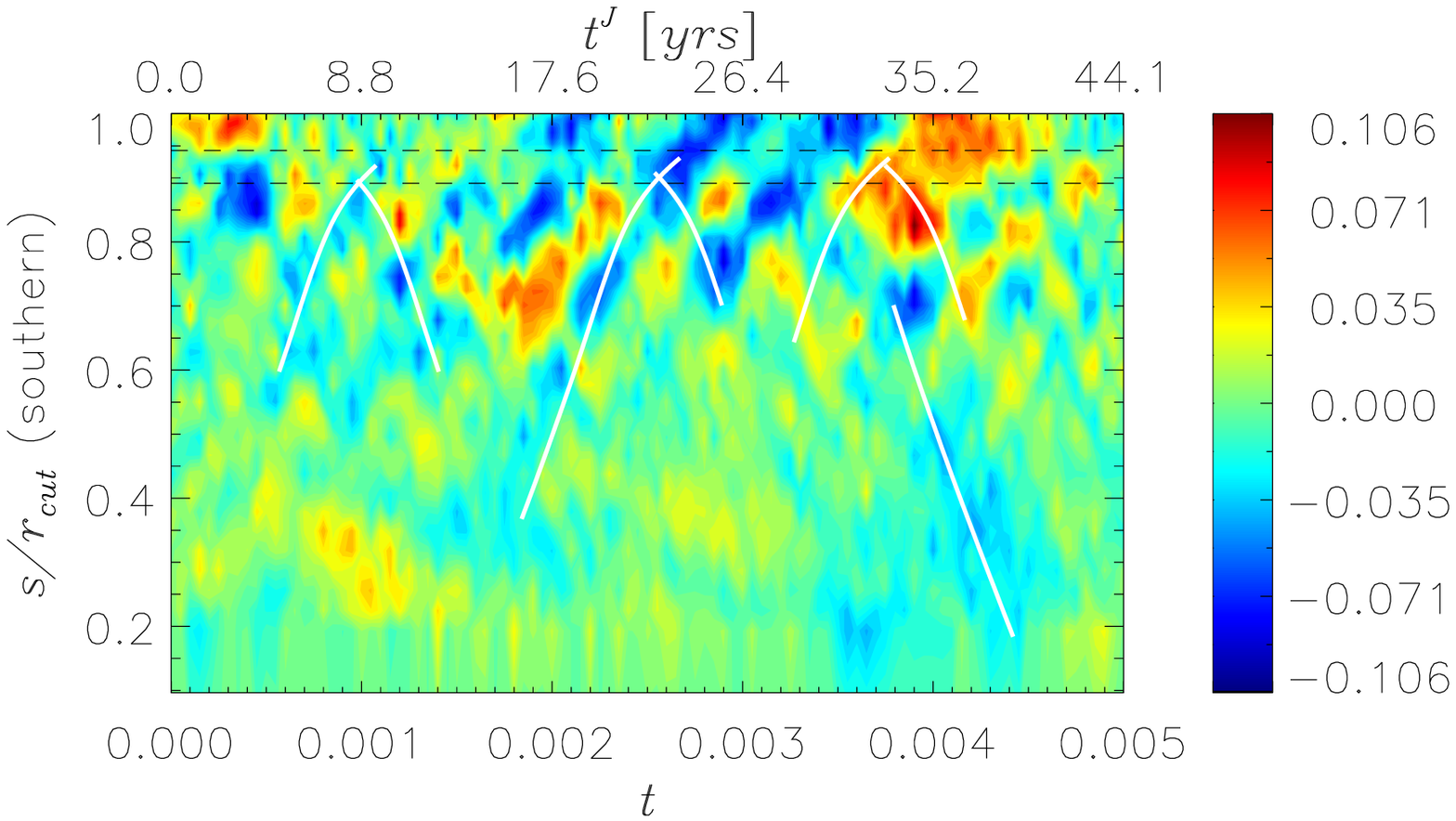}&
 \includegraphics[width=75mm]{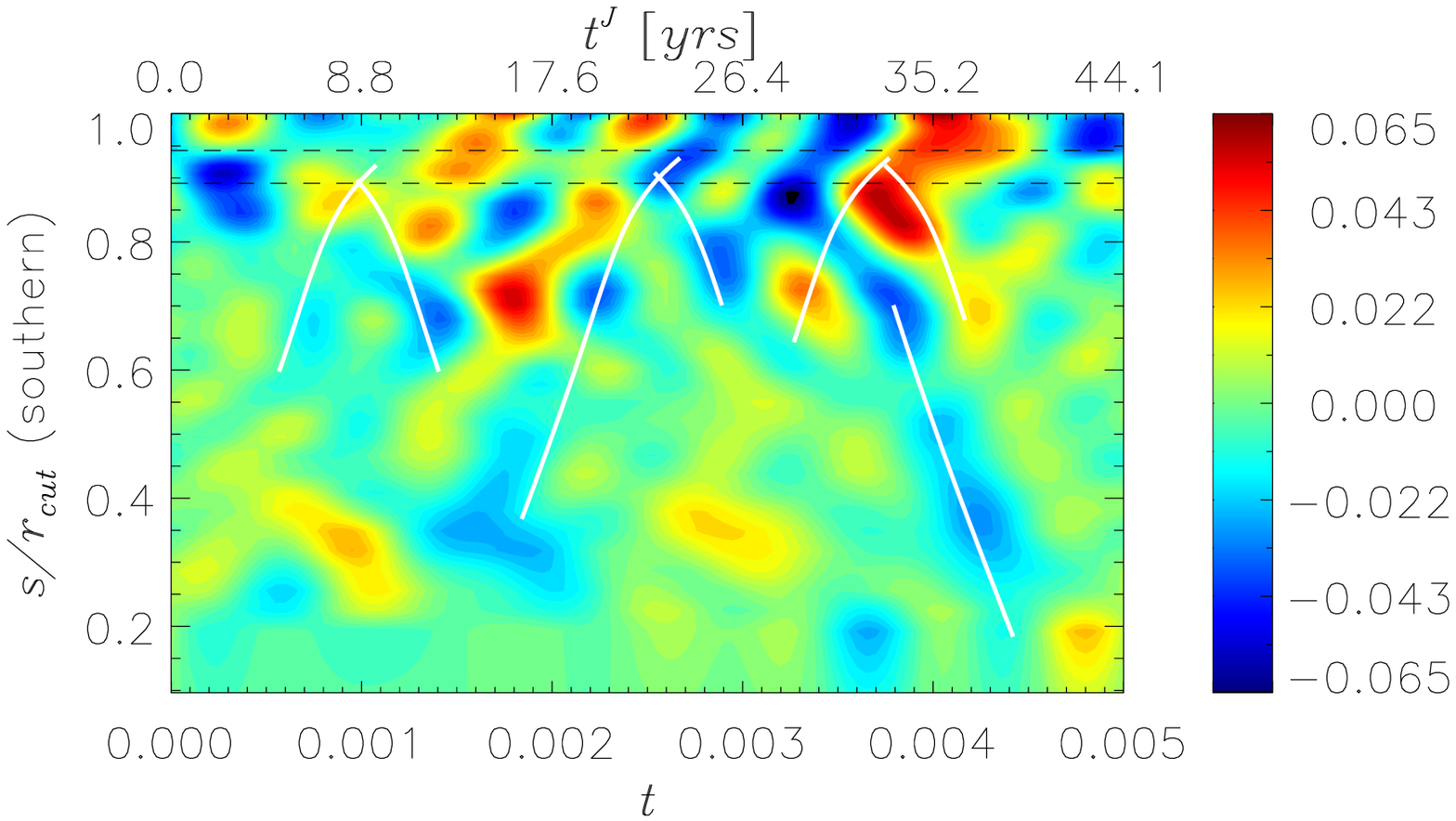}\\
 \end{tabular}
 \caption{(a-b) Zonal flow fluctuations, $\overline{u'_\phi}$, on the surface of the cut-off boundary, $r_\tx{cut} \sim 0.96\,R_\tx{J}$, for run E.
 In northern (a) and southern (b) hemispheres.
 The ordinates represent the cylindrical radius ${s}/{r_\tx{cut}}$
 so that white curves indicate the same phase paths as those shown
 in fig.~\ref{fig:axisymmetic_part}b.
 \revone{The horizontal dashed lines denote the range of the MTC radii,
 ${s}/{R_\tx{J}} \sim 0.85$ and $0.90$,
 corresponding to latitudes $\sim 32^\circ$ and $26^\circ$ on the surface, respectively.}
 The amplitude is scaled by the maximum of the steady part,
 $\overline{\widetilde{u_\phi}}$,
 corresponding to about 1.4\,m\,s${^{-1}}$ in the dimensional unit;
 see table \ref{table:dimensional_outputs}.
 (c-d) Same as figures a-b but with periods outside 0.00063$\leq t \leq$0.0025, 
 or 5.6 yrs $\leq t^\tx{J} \leq$ 22 yrs, filtered from the data.}
 \label{fig:surface_flow_fluctuations}
\end{figure}

%%%%%%%%%%%%%%%%%%%%%%%%%%%%%%%%%%%%%
%%%%%%%%%%%%%%%%%%%%%%%%%%% appendices
\clearpage
%\nolinenumbers
\appendix

\revtwo{
\section{Anelastic TWs with the momentum disturbances} \label{sec:momentum}
}

When choosing to formulate anelastic TWs for the momentum,
 $\langle \rho_\tx{eq} \overline{u'_\phi} \rangle$,
 the wave equation may be given by
\begin{equation}
\label{waveeq_momentum}
 \frac{\partial^2}{\partial t^2} \frac{\langle \rho_\tx{eq} \overline{u'_\phi} \rangle}{s}
 = \frac{1}{s^3 h} \frac{\partial}{\partial s}
     \left( 
       s^3 h  \; {\hat{U}_\tx{A}}^2
       \frac{\partial}{\partial s} \frac{\langle \rho_\tx{eq} \overline{u'_\phi} \rangle}{s}
     \right)  \;,
\end{equation}
instead of eq.~(\ref{waveeq}).
Here ${\hat{U}_\tx{A}^2} = {\langle \overline{\widetilde{B_s^2}}/{\mu_0 \rho_\tx{eq}} \rangle}$.
Consequently the axial angular momentum change in the metallic region
 is calculated through  
 \begin{equation}
  \delta \hat{\sigma} = 2\pi \int_{s_\tx{tc}}^{s_\tx{mtc}}
   \int_{z_{-}}^{z_{+}} s^2 h \langle {\rho}_\tx{eq} \overline{u'_\phi} \rangle\,dz\,ds 
 \end{equation}
 (cf. eq.~\ref{eq:delta_sigma}).

Profiles of the Alfv\'{e}n speeds $\hat{U}_\tx{A}$ in our simulations 
 are very similar to those for the earlier formulation,
 so we avoid presenting these plots.
In table \ref{table:dynamo_simulations__momentum}
 we examine the wave speeds,
 the resulting traveltimes $\hat{\tau}_\tx{A}$ across the metallic region,
 and the maximal amplitudes of the angular momentum $\delta \hat{\sigma}$
 for the present formulation. 
Compared to values listed in table \ref{table:dynamo_simulations},
 the speed $\hat{U}_\tx{A}$ at $s_\tx{mtc}$ increases by 7-9 \% and
 the traveltime $\hat{\tau}_\tx{A}$ gets shorter by 4-7 \%.
The influence on $\delta \hat{\sigma}$ is within 7 \%. 

Figure \ref{fig:axisymmetic_part_momentum} depicts contours of 
 ${\langle \rho_\tx{eq} \overline{u'_\phi} \rangle}$ 
 in $s$-$t$ space for model E.
As the density diminishes in the weakly conducting zone, 
 the momentum plot does not exhibit the features seen outside the MTC
 in figure \ref{fig:axisymmetic_part}b
 but highlights disturbances at small $s$.
The phase paths calculated with $\hat{U}_\tx{A}$ account for the patterns
 in the model.

%\vspace{50mm}
%%%%%%%%%%%%%%%%%%%%%%%%%%% tables
\clearpage
\begin{table}[th]
\setcounter{table}{0}
 \caption{Dimensionless output parameters of the simulations
 for the current formulation (cf. table~\ref{table:dynamo_simulations}).
 The columns ${\hat{U}_A (s_\tx{mtc})}$, $\hat{\tau}_\tx{A}$, and $\delta \hat{\sigma}$
 represent, respectively, 
 the Alfv\'{e}n speed at the MTC,
 the traveltime across the conducting region
 (from the core boundary to the MTC),
 and the maximal amplitude of the axial angular momentum fluctuation in the
 metallic region.}
\label{table:dynamo_simulations__momentum}
\centering\footnotesize
\vspace{3mm}
\begin{tabular}{l lll}
\hline
 Run  & $\hat{U}_\tx{A}(s_\tx{mtc})$ & $\hat{\tau}_\tx{A}$ & $\delta \hat{\sigma}$  \\
\hline
 A & 316 & 0.0015 & 38.6  \\
 E & 452 & 0.00116 & 41.45 \\
 I & 373 & 0.00111 & 41.43 \\
\hline
\end{tabular}
%\tablenotetext{a}{Footnote text here.}
\end{table}

%%%%%%%%%%%%%%%%%%%%%%%%%%
\vspace{10mm}

\begin{figure}[bh]
\setcounter{figure}{0}
 \includegraphics[width=95mm]{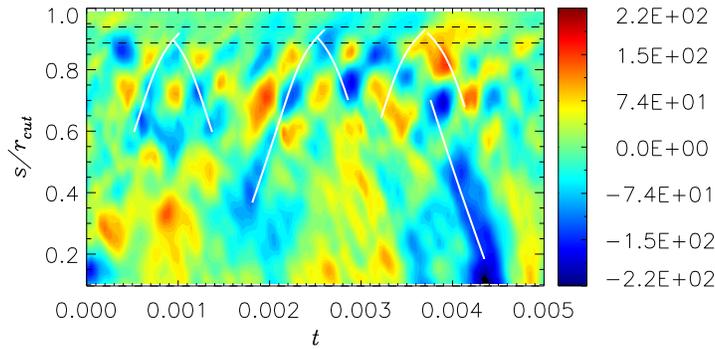}
 \caption{The fluctuating, $z$-averaged azimuthal momentum,
 $\langle \rho_\tx{eq} \overline{u'_\phi} \rangle$,
 for run E [cf. fig.~\ref{fig:axisymmetic_part}b].
 White curves indicate phase paths of the Alfv\'{e}n speed, $\hat{U}_\tx{A} = \langle \overline{\widetilde{B_s^2}}/{\mu_0 \rho_\tx{eq}} \rangle$, in the present formulation.
 The horizontal dashed lines indicate a range of the MTC radius,
 ${s}/{r_\tx{cut}} \sim 0.89$ and $0.94$.
 Note that the amplitude $\langle \rho_\tx{eq} \overline{u'_\phi} \rangle$
 is smaller outside the MTC than in the metallic interior,
 whereas $\langle \overline{u'_\phi} \rangle$ is substantial outside the MTC
 (see fig.\ref{fig:axisymmetic_part}b).}
 \label{fig:axisymmetic_part_momentum}
\end{figure}

%%%%%%%%%%%%%%%%%%%%%%%%%%

\clearpage
\revone{
\section{Spectral analysis of the internal flow fluctuations} \label{sec:axisymmetic_part_filtered}
}

Though the previous plots
 have exhibited the characteristics of waves,
 filtering over the data highlights their signals more clearly.
Figure \ref{fig:axisymmetic_part_filtered}a shows
 $\langle \overline{u'_\phi} \rangle$ for model A,
 removing modes outside the period range $\tau = [0.00031$ - $0.0025]$ by Fourier transformation 
 and can be compared with the full data from fig.~\ref{fig:axisymmetic_part}a.
Some intermittent standing wave features near the MTC are more noticeable here. 
Similarly, figure \ref{fig:axisymmetic_part_filtered}b excludes
 periods outside the range $\tau = [0.00063$ - $0.0025]$ for model E and
 better illustrates the oscillation and propagation
 seen in fig.~\ref{fig:axisymmetic_part}b.
In figure \ref{fig:axisymmetic_part_filtered}c for model I, 
 a period range $\tau = [0.00031$ - $0.0013]$ is used for the transformation.
Note that the time series is now extended - with both earlier and later times displayed - in this plot compared to fig.~\ref{fig:axisymmetic_part}c.
The spectral analysis here leaves clean travelling features, rather than reflecting and/or standing waves.

%%%%%%%%%%%%%%%%%%%%%%%%%%
\begin{figure}
\setcounter{figure}{0}
 \revone{(a)} \\
 \includegraphics[width=95mm]{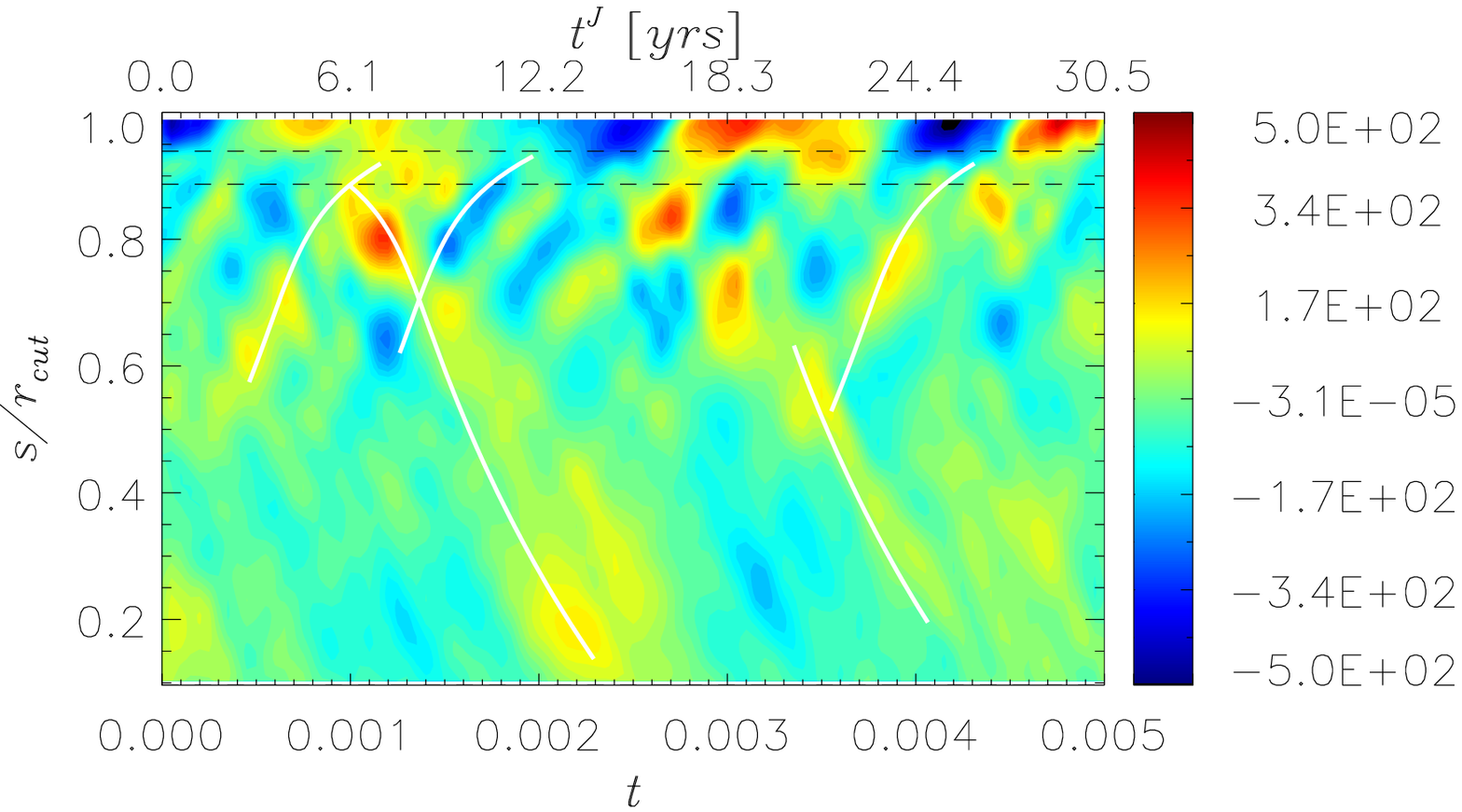} \\
 \revone{(b)} \\
 \includegraphics[width=95mm]{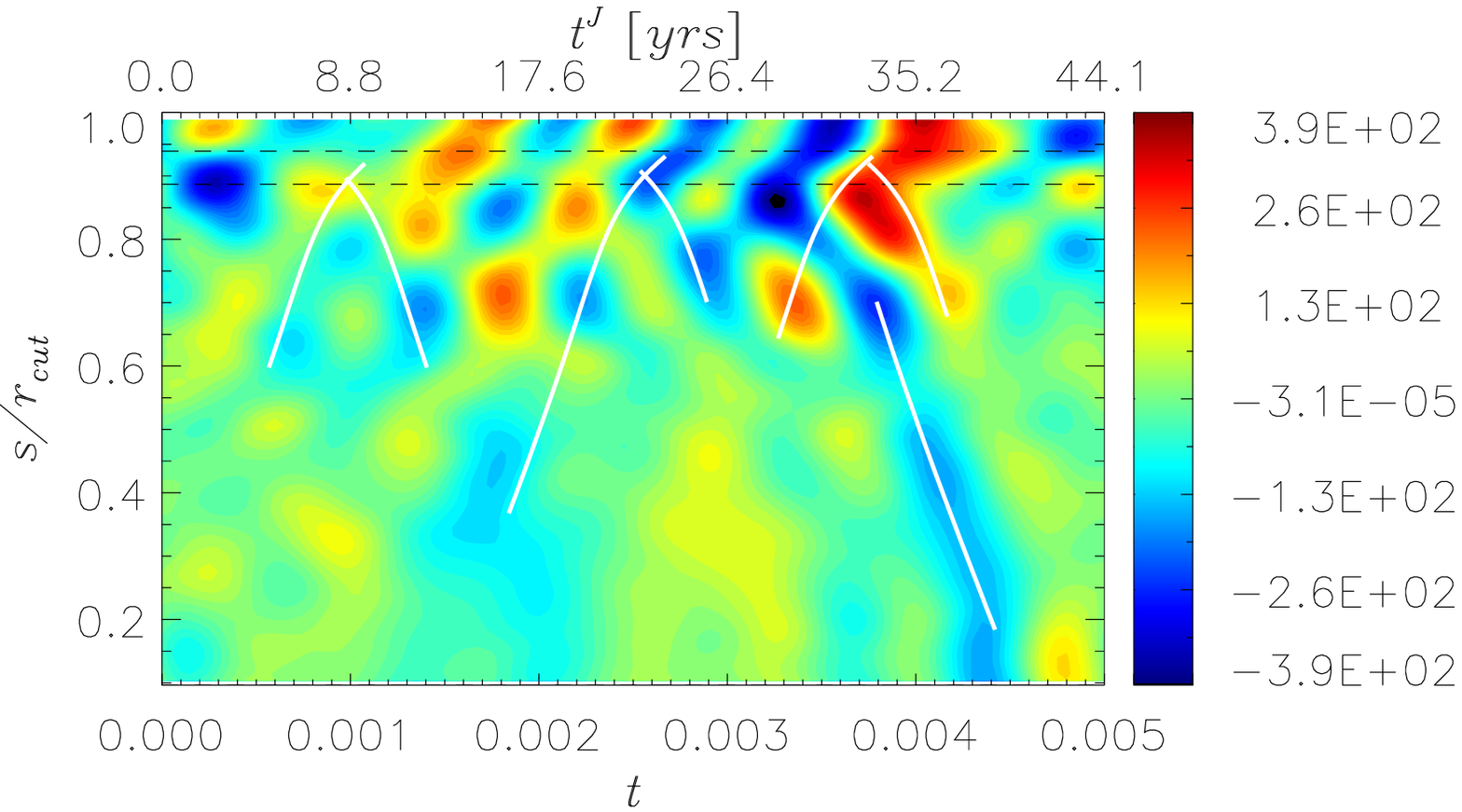} \\
 \revone{(c)} \\
 \includegraphics[width=95mm]{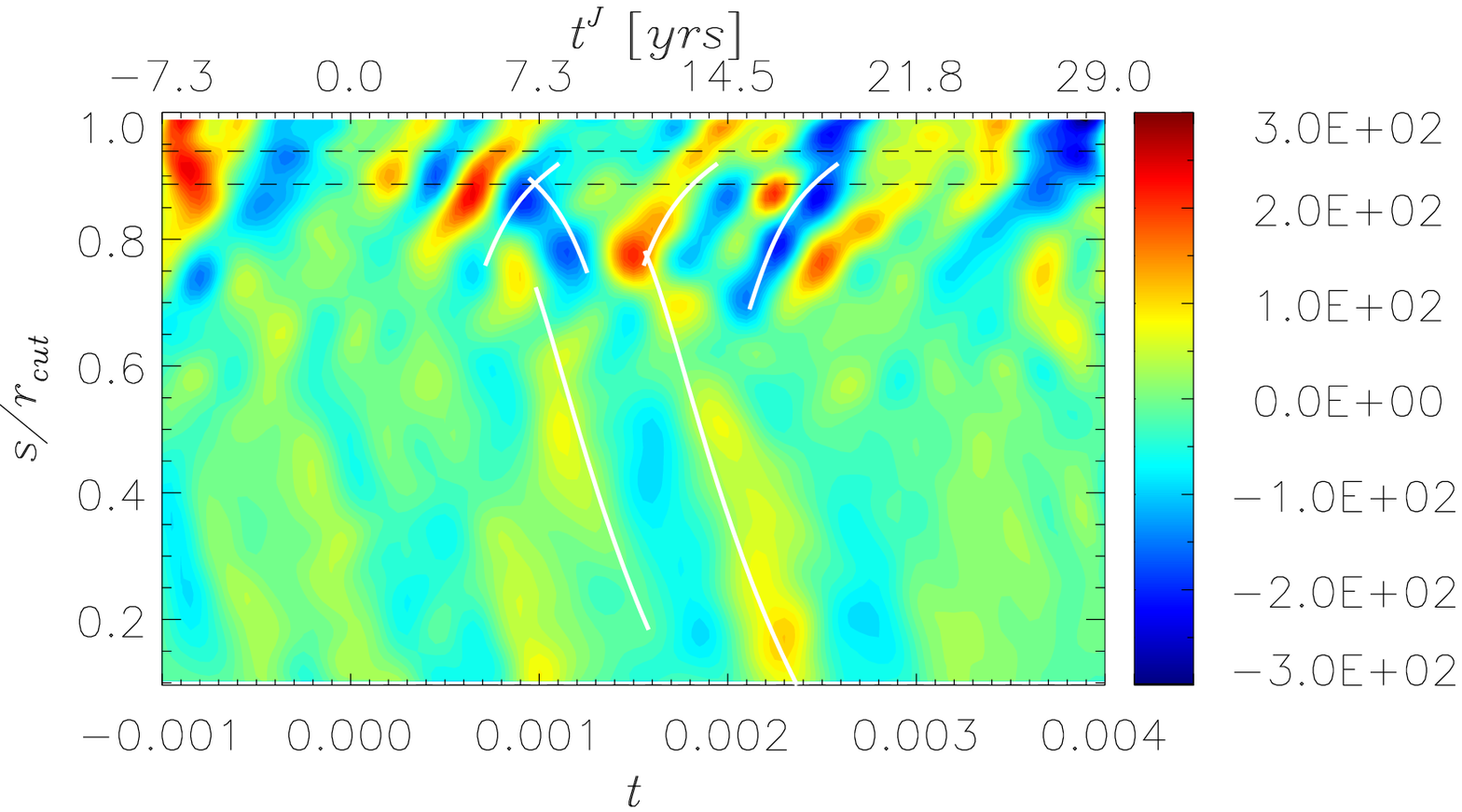} 
 \vspace{-2mm}
 \caption{The $z$-averaged azimuthal velocity,
 $\langle \overline{u'_\phi} \rangle$, in which
 all the periods are filtered out except 0.00031$\leq t \leq$0.0025 in run A (a), 
 0.00063$\leq t\leq$0.0025 in run E (b),
 and 0.00031$\leq t\leq$0.0013 in run I (c).
 Note that figure c displays the filtered version of the data from 
 fig.~\ref{fig:axisymmetic_part}c, but the time series shown here is longer.
 \revone{The horizontal dashed lines indicate the range of the MTC radii.}}
 \label{fig:axisymmetic_part_filtered}
\end{figure}
%%%%%%%%%%%%%%%%%%%%%%%%%%

\clearpage 
\revone{
\section{Alfv\'{e}n waves approaching a resistive zone} \label{sec:appendix_1d}
}

We consider a Cartesian, one-dimensional model for Alfv\'{e}n waves approaching a resistive layer.
Let $x=0$ be the interface between a perfectly conducting fluid (for negative $x$)
 and a weakly conducting one
 (for positive $x$). 
They are permeated by a uniform background magnetic field
 $B_0$ in the $x$ direction. 
For simplicity we assume an incompressible fluid
 with $\rho_0$ being constant density.
We then suppose the variables  
\begin{equation}
 \mib{B} = B_0 \hat{\mib{e}}_x + b_y (x) \hat{\mib{e}}_y \; , \quad 
 \mib{u} = u_y (x) \hat{\mib{e}}_y 
\end{equation}
to rewrite the equations of induction and momentum as 
\begin{eqnarray}
 \frac{\partial b_y}{\partial t}
 = B_0 \frac{\partial u_y}{\partial x}
  + \frac{\partial}{\partial x} 
      \eta \frac{\partial b_y}{\partial x} \qquad \textrm{and} \qquad
 \frac{\partial u_y}{\partial t}
 = \frac{B_0}{\mu_0 \rho_0} \frac{\partial b_y}{\partial x} \; ,
   \label{eq:induc_momentum}
\end{eqnarray}
respectively.
Here the magnetic diffusivity, $\eta (x) = {1}/{\mu_0\,\sigma(x)}$,
 varies in $x$: it is set zero for $x < 0$
 and to a constant nonzero value $\eta_0$ for $x > 0$. 
At the interface the field and velocity are continuous,
 i.e. the continuity condition across $x =0$ is required
 for $b_y$ and ${\partial b_y}/{\partial x}$.

Eq. (\ref{eq:induc_momentum}) may be reduced to 
\begin{eqnarray}
 \frac{\partial^2 b_y}{\partial t^2}
 &=& V_\tx{A}^2 \frac{\partial^2 b_y}{\partial x^2}
       \qquad \textrm{for} \quad x < 0     \\ 
 \frac{\partial^2 b_y}{\partial t^2}
 &=& V_\tx{A}^2 \frac{\partial^2 b_y}{\partial x^2} 
  + \eta_0 \frac{\partial^3 b_y}{\partial t \partial x^2} 
       \qquad  \textrm{for} \quad x > 0 \; ,    \label{eq:wave_eq}
\end{eqnarray}
where the Alfv\'{e}n speed $V_\tx{A} = {B_0}/{\sqrt{\rho_0 \mu_0}}$.
Now we seek solutions of the form 
\begin{eqnarray}
 b_y &=& e^{\mathrm{i}\omega t} \left( e^{-\mathrm{i}kx} + \mathcal{R} e^{\mathrm{i}kx}  \right)
  \qquad \textrm{for} \quad x < 0 \\
 b_y &=& \mathcal{T} e^{\mathrm{i}\omega t}  e^{\lambda x}
  \qquad \textrm{for} \quad x > 0   \label{eq:solution_negative_x}
\end{eqnarray}
where $\lambda$, $\mathcal{R}$ and $\mathcal{T}$ are complex and
 $k^2 = {\omega^2}/{V_\tx{A}^2}$.
For $x>0$, substituting (\ref{eq:solution_negative_x}) into the respective wave equation (\ref{eq:wave_eq}) gives 
\begin{equation}
 \lambda^2
 = -\frac{\omega^2 (V_\tx{A}^2 - \mathrm{i} \omega \eta_0)}
         {V_\tx{A}^4 + \omega^2 \eta_0^2}   \; .
\end{equation}
When the waves travel quickly so that $\omega {\eta_0} \gg {V_\tx{A}^2}$, 
the valid solution is 
\begin{equation}
 \lambda = - (1 + \mathrm{i}) \sqrt{\frac{\omega}{2\eta_0}} \; .
\end{equation}
Notice here the electromagnetic skin depth given with $\sqrt{2 \eta_0/\omega}$.
So the continuity condition on $b_y$ and ${\partial b_y}/{\partial x}$
 implies, respectively,
 \begin{equation}
  1 + \mathcal{R} = \mathcal{T} \qquad \textrm{and} \qquad  
  \mathrm{i}k (-1 + \mathcal{R})
  = - (1+\mathrm{i}) \sqrt{\frac{\omega}{2 \eta_0}} \; \mathcal{T}  \; .
 \end{equation}
We hence obtain the reflection coefficient,
\begin{equation}
 \mathcal{R} = \frac{\mathrm{i}k - (1+\mathrm{i}) \sqrt{\omega/2\eta_0}}
                    {\mathrm{i}k + (1+\mathrm{i}) \sqrt{\omega/2\eta_0}}  \; .
\end{equation}
For $\omega/{\eta_0} \gg k^2$, 
 this yields $\mathcal{R} \rightarrow -1$ and $\mathcal{T} \rightarrow 0$,
 i.e. nearly perfect reflection. 
From (\ref{eq:induc_momentum}),
 this is equivalent to positive reflection of $u_y$ across the interface. 
When the approximations are inappropriate,
 it gives rise to partial reflection and partial transmission.

\end{document}